\documentstyle[aps,preprint,eqsecnum,epsf]{revtex}
\newtheorem{lem}{Lemma}
\newcommand{\postscript}[2]
   {\setlength{\epsfxsize}{#2\hsize}
   \centerline{\epsfbox{#1}}}
\begin{document}
\baselineskip = 0.54cm
\parbox{0.8\textwidth}
{{\Large\bf Higher-order solitons in the
$N$-wave system }}\\
\begin{center}
\parbox{0.8\textwidth}{{\sf  Valery S. Shchesnovich}${}^{a)}$ {\sf and
Jianke Yang}${}^{b)}$\\
{\it Department of Mathematics and Statistics,
University of Vermont, Burlington VT 05401, USA}\\[0.5cm]

\baselineskip=0.54cm

The soliton dressing matrices for the  higher-order  zeros of the
Riemann-Hilbert problem for the $N$-wave system are considered. For the
elementary higher-order zero, i.e. whose algebraic multiplicity is arbitrary
but the geometric multiplicity is 1, the general soliton dressing matrix is
derived. The theory is applied to the study of higher-order soliton solutions
in the three-wave interaction model. The simplest higher-order soliton solution
is presented. In the generic case, this solution describes the breakup of a
higher-order pumping wave into two higher-order elementary waves, and the
reverse process. In non-generic cases, this solution could describe ($i$) the
merger of a pumping sech wave and an elementary sech wave into two elementary
waves (one sech and the other one higher-order); ($ii)$ the breakup of a
higher-order pumping wave into two elementary sech waves and one pumping sech
wave; and the reverse processes. This solution could also reproduce fundamental
soliton solutions as a special case. }
\end{center}

\vspace{1cm} Keywords: $N$-wave interaction; higher-order solitons;
Riemann-Hilbert problem.

\vfill \hrule
\medskip

\noindent
{{\small\rm  ${}^{a)}$  Current address: Instituto de F{i}sica Te\'{o}rica, Universidade Estadual Paulista,
 Rua Pamplona 145, 01405-900 S\~{a}o Paulo, Brazil\\
 Email: valery@ift.unesp.br}
\medskip

\noindent
{\small\rm ${}^{b)}$ Email: jyang@emba.uvm.edu}
\newpage

\section{Introduction}

Soliton solutions to nonlinear partial differential equations  (PDEs) in (1+1)
dimensions have been of great interest ever since the very discovery of
completely integrable PDEs. There is a wide range of literature concerning
integrable nonlinear PDEs and their soliton solutions (see, for instance,
Refs.~\cite{AS81,NMPZ84,FT87,AC91} and the references therein). Much is known
about the behavior of solitons and their interactions in various integrable
systems (soliton scattering, breather solutions, soliton bound states, etc).
Such knowledge is very valuable not only for the underlying integrable systems,
but also for nearly integrable systems which can be studied analytically by
soliton perturbation theories.

It is an astonishing fact, however, that notwithstanding the almost  three
decades of advances in the study of soliton dynamics, still there are
substantial gaps in our knowledge of soliton solutions in integrable nonlinear
PDEs. Indeed, the reader familiar with the basics of the inverse scattering
transform method knows that it is poles of the reflection coefficient (or, in
modern terms, zeros of the Riemann-Hilbert problem) that give rise to the
soliton solutions. The soliton solutions are usually derived by using one of
the several well-known techniques, such as the dressing method
\cite{AS81,ZS79a,ZS79b} or the Riemann-Hilbert problem approach
\cite{NMPZ84,FT87}. However, in most publications (except, to our knowledge,
Refs.~\cite{ZS72,Wadati,Tsuru,OptLett,Nathalie,ablowitz,ablowitz2}, which are
discussed below) only soliton solutions from simple poles are considered. It is
usually assumed that a multiple-pole solution can be obtained in a
straightforward way by coalescing several distinct poles (see, for instance,
Ref.~\cite{NMPZ84,Kawata}) which describe multisoliton solutions. This would be
indeed the  case if such coalescing were a {\em regular limit}. However, this
limit is obviously a singular one. Indeed, the soliton dressing matrix
corresponding to a multisoliton solution is a rational matrix function which
has distinct {\em simple} poles, while the coalescing procedure must produce
{\em multiple} poles. Obviously, a more careful examination of this issue is
necessary. This is the main subject of the present paper.

Soliton solutions corresponding to multiple poles, i.e., the higher-order
solitons, have been investigated in the literature before. A soliton solution
to the nonlinear Schr\"odinger (NLS) equation corresponding to a double pole
was first given in Ref.~\cite{ZS72} but without much analysis. The double- and
triple-pole soliton solutions to the KdV equation were examined in
Ref.~\cite{Wadati} and the general $N$-pole soliton solution to the sine-Gordon
equation was extensively studied in Ref.~\cite{Tsuru} using the associated
Gelfand-Levitan-Marchenko equation. In Refs.~\cite{OptLett,Nathalie},
higher-order soliton solutions to the NLS equation were studied by employing
the dressing method. Finally, in \cite{ablowitz,ablowitz2}, higher order
solitons in the Kadomtsev-Petviashvili~I equation were derived by the inverse
scattering method.

In this article, we consider the higher-order zeros of the Riemann-Hilbert
problem for the $N$-wave system and study the corresponding soliton matrices.
We derive the soliton dressing matrices for the simplest class of higher-order
zeros -- the elementary higher-order zeros. We call the $n$-th order zero $k_1$
of the Riemann-Hilbert problem elementary if  the soliton matrix evaluated at
$k_1$ has only one vector in the kernel, i.e. the geometric multiplicity of the
zero is 1 (see Definition 1 in Sec. 3 for more details). The corresponding
higher-order soliton solutions to the $N$-wave system are determined. Then we
apply our theory to the physically important three-wave interaction model and
derive the simplest higher-order soliton solution. In the generic case, this
solution describes the breakup of a higher-order pumping wave into two
higher-order elementary waves, and the reverse process. In non-generic cases,
this solution could describe ($i$) the merger of a pumping sech wave and an
elementary sech wave into two elementary waves (one sech and the other one
higher-order); ($ii$) the breakup of a higher-order pumping wave into three
sech waves --- one pumping wave and two elementary waves. The higher-order
soliton solution could also reproduce fundamental soliton solutions as a
special case.

In general, one needs to consider zeros with the geometric multiplicities
taking values from 1 to $N-1$, where $N$ is the matrix dimension of the
Riemann-Hilbert problem. The present work is the first step towards the solution of this
general case. The point is that the soliton matrices derived here for the
elementary zeros provide the building blocks for the most general case. We plan
to address the general problem in the next paper. Thus, for the 3-wave
interaction model, the higher-order soliton solutions we derived in this paper
(which correspond to elementary higher-order zeros) may not be the most general
higher-order soliton solutions in this wave system.

 It is noted that the soliton dressing matrices for the higher-order soliton
solutions were already a subject of interest in Refs.~\cite{OptLett,Nathalie},
where an ansatz for the soliton matrices was proposed for the $2\times2$
Zakharov-Shabat spectral problem and expressions for the higher-order soliton
solutions of the nonlinear Schr\"odinger equation were obtained. In the present
paper, we study higher-order solitons in the $N$-wave system. Even though the
evolution equations considered in \cite{OptLett,Nathalie} and this article are
different, their Riemann-Hilbert formulations have a lot in common. In this
paper, we place the emphasis on the {\em derivation} of the soliton matrices
for the higher-order zeros and fill some gaps in the approach of
Ref.~\cite{Nathalie}. We show that the ansatz of \cite{Nathalie} is precisely
the soliton matrix for an {\em elementary} higher-order zero in a $N\times N$
spectral problem. For the $2\times 2$ Zakharov-Shabat spectral problem, any
higher-order zero is elementary. Thus the ansatz of \cite{Nathalie} is the
general soliton matrix of the Zakharov-Shabat problem. Consequently,
higher-order soliton solutions obtained in \cite{OptLett,Nathalie} are the
general higher-order soliton solutions in the nonlinear Schr\"odinger equation.
But in a $N\times N$ spectral problem with $N>2$, a higher-order zero is
non-elementary in general. In that case, the ansatz of \cite{Nathalie} will not
be the general form of soliton matrices. In this paper, we also uncover the
invariance property of the soliton matrix for a higher-order zero, which is
necessary for the  vector-parametrization of the soliton matrix to be
self-consistent. The invariance property obtained in \cite{Nathalie} is shown
to be just a special case. Lastly, we point out that our derivation of
higher-order solitons in the $N$-wave system is made for general dispersion
laws and arbitrary matrix dimensions. In addition, our results can be
generalized to the most general case of non-elementary zeros of the $N\times N$
spectral problem.

The paper is organized as follows. A summary on the Riemann-Hilbert   problem
is placed in section \ref{secRH}. Section \ref{multpl} is the central section
of the paper. There we present the theory of soliton matrices corresponding to
the higher-order zeros of the Riemann-Hilbert problem. For an elementary
higher-order zero of the $N\times N$ Riemann-Hilbert problem (see definition in
the text), the general soliton matrix is derived. This soliton matrix is
similar to the ansatz as proposed in \cite{Nathalie}. In section \ref{applica},
the theory is applied to derive the simplest higher-order soliton solution in
the three-wave interaction model.

\section{The Riemann-Hilbert problem approach: summary}
\label{secRH}

The integrable nonlinear PDEs in 1+1 dimensions are associated  with the matrix
Riemann-Hilbert problem (consult, for instance, Refs.
\cite{AS81,NMPZ84,FT87,AC91,ZS79a,ZS79b,Fokas1,Fokas2,Fokas3,Leon,BC1,BC2,BC3,BDT,Zhou1,Zhou2}).
The matrix Riemann-Hilbert problem (below we work in the space of $N\times N$
matrices) is the problem of finding the holomorphic factorization, denoted
below by  $\Phi_+(k)$ and $\Phi^{-1}_-(k)$, in the complex plane of  a
nondegenerate matrix function $G(k)$ given on an oriented curve~$\gamma$:
\begin{eqnarray}
\Phi_-^{-1}(k,x,t)\Phi_+(k,x,t) &=& G(k,x,t)\nonumber\\
&\equiv&\exp{[-\Lambda(k)x-\Omega(k)t]}
G(k,0,0)\exp{[\Lambda(k)x+\Omega(k)t]},\quad k\in\gamma.
\label{RH1}\end{eqnarray}
Here the matrix functions $\Phi_+(k)$ and $\Phi_-^{-1}(k)$ are holomorphic in
the two complementary domains of the complex $k$-plane: $C_+$ to the left and
$C_-$ to the right from the curve $\gamma$, respectively. The matrices
$\Lambda(k)$ and $\Omega(k)$ are called the dispersion laws. In this paper, we
require the dispersion laws to be diagonal (we have accounted for this on the
right-hand side of equation (\ref{RH1}) by writing the explicit
$(x,t)$-dependence in the form of an exponent). The Riemann-Hilbert problem
requires an appropriate normalization condition. Usually the curve $\gamma$
contains the infinite point $k=\infty$ of the complex plane and the
normalization condition is formulated as
\begin{equation}
\Phi_\pm(k,x,t) \to I,\quad \text{as}\quad k\to \infty.
\label{RH2}\end{equation}
This normalization condition is called the canonical normalization.
Setting the normalization condition to an arbitrary nondegenerate
matrix function $S(x,t)$ leads to the gauge equivalent integrable
nonlinear PDE, e.g., the Landau-Lifshitz equation in the case of the
NLS equation \cite{FT87}. Obviously, the new solution
$\hat{\Phi}_\pm(k,x,t)$  to the Riemann-Hilbert problem, normalized to
$S(x,t)$, is related to the canonical solution by the following
transformation
\begin{equation}
\hat{\Phi}_\pm(k,x,t) = S(x,t)\Phi(k,x,t).
\label{gaugePhi}\end{equation}
Thus, without any loss of generality, we confine ourselves to the
Riemann-Hilbert problem under the canonical normalization.

For physically applicable nonlinear PDEs the Riemann-Hilbert problem possesses
the involution properties, which reduce the number of the dependent variables
(complex fields). The $N$-wave interaction model admits the following
involution property of the associated Riemann-Hilbert problem
\begin{equation}
\Phi_+^\dag(k) = \Phi_-^{-1}(\overline{k}),\quad \overline{k} \equiv
k^*.\label{invol}\end{equation} Here the superscript ``$\dag$''
represents the Hermitian conjugate, and ``*'' the complex conjugate.
However, our approach can be trivially extended to the general case
without such involution. To keep our treatment general, we will use the
overlined quantities where applicable. The reduction to the involution
is then done by associating the overline with the Hermitian conjugation
in the case of vectors and matrices and with the complex conjugation in
the case of scalar quantities.

To solve the Cauchy problem for the integrable nonlinear PDE posed on
the whole axis $x$, one usually constructs the associated
Riemann-Hilbert problem  starting with the linear spectral equation
\begin{equation}
\partial_x \Phi(k,x,t) = \Phi(k,x,t)\Lambda(k) + U(k,x,t)\Phi(k,x,t),
\label{RH3}\end{equation}
whereas the $t$-dependence is given by a similar equation
\begin{equation}
\partial_t \Phi(k,x,t) = \Phi(k,x,t)\Omega(k) + V(k,x,t)\Phi(k,x,t).
\label{RH4}\end{equation}
The nonlinear integrable PDE corresponds to the compatibility condition
of the system (\ref{RH3}) and (\ref{RH4}):
\begin{equation}
\partial_tU-\partial_xV+[U,V]=0.
\label{RH5}\end{equation}

The essence of the approach based on the Riemann-Hilbert problem lies
in the fact that the evolution governed by the complicated nonlinear
PDE (\ref{RH5}) is mapped to the evolution of the spectral data given
by simpler equations such as (\ref{RH1}) and (\ref{RH8}). For details,
consult Refs.~\cite{FT87,AC91,Fokas1,Fokas2,Fokas3,Leon}.

Let the evolution equations for the spectral data be given. In our
case, these are equation (\ref{RH1}) for $G$ and equation (\ref{RH8})
(see below) for the discrete data.  Then the matrices $U(k,x,t)$ and
$V(k,x,t)$ describing the evolution of $\Phi_\pm$ can be retrieved from
the Riemann-Hilbert problem. In our case, the potentials $U(k,x,t)$ and
$V(k,x,t)$ are completely determined  by the (diagonal) dispersion laws
$\Lambda(k)$ and $\Omega(k)$  and the Riemann-Hilbert solution
$\Phi\equiv\Phi_\pm(k,x,t)$. Indeed, let us assume that the dispersion
laws are polynomial functions, i.e.,
\begin{equation}
\Lambda(k) = \sum_{j=0}^{J_1}A_jk^j,\quad \Omega(k)=\sum_{j=0}^{J_2}
B_j k^j.
\label{polynom}\end{equation}
Then using similar arguments as in Ref.~\cite{Leon} we get:
\begin{equation}
U = -{\cal P}\{\Phi\Lambda\Phi^{-1}\}, \quad V = -{\cal
P}\{\Phi\Omega\Phi^{-1}\}.
\label{UV}\end{equation}
Here the matrix function $\Phi(k)$ is expanded into the asymptotic
series,
\[
\Phi(k)=I+k^{-1}\Phi^{(1)}+k^{-2}\Phi^{(2)}+..., \quad k\to \infty,
\]
and the operator ${\cal P}$ cuts out the polynomial asymptotics of its
argument as $k\to\infty$. An important property of matrices $U$ and $V$
is that
\begin{equation}
\text{Tr}U(k,x,t) = - \text{Tr}\Lambda(k),\quad \text{Tr}V(k,x,t) = -
\text{Tr}\Omega(k),
\label{Traces}\end{equation}
which evidently follows from equation (\ref{UV}). Below, let us
consider the three-wave interaction system as an example
\cite{NMPZ84,3wave1,3wave2,3wave3}. Set $N=3$,
\begin{equation}
\Lambda(k) = ikA,\quad A =  \left(\begin{array}{ccc}a_1 & 0 & 0\\
0 & a_2 & 0\\ 0 & 0 & a_3  \end{array}\right),\qquad
\Omega(k) = ikB,\quad B =  \left(\begin{array}{ccc}b_1 & 0 & 0\\
0 & b_2 & 0\\ 0 & 0 & b_3  \end{array}\right),
\label{disper}\end{equation}
where $a_j$ and $b_j$ are real with the elements of $A$  being ordered:
$a_1>a_2>a_3$. From equation (\ref{UV}) we get
\begin{equation}
U = -\Lambda(k) + i[A,\Phi^{(1)}],\quad V = -\Omega(k) +
i[B,\Phi^{(1)}].
\label{UV3wave}\end{equation}
Setting
\begin{equation}
u_1 = \sqrt{a_1-a_2}\Phi^{(1)}_{12},\quad u_2 =
\sqrt{a_2-a_3}\Phi^{(1)}_{23},\quad u_3 =
\sqrt{a_1-a_3}\Phi^{(1)}_{13},
\label{u1u2u3}\end{equation}
assuming the involution (\ref{invol}),  and using equation (\ref{UV3wave}) in
(\ref{RH5}) we get the three-wave system:
\begin{mathletters}
\label{3wave}
\begin{eqnarray}
\partial_t u_1 + v_1 \partial_x u_1 +i\varepsilon\overline{u}_2u_3 &=&
0,\\
\partial_t u_2 + v_2 \partial_x u_2 +i\varepsilon\overline{u}_1u_3 &=&
0,\\
\partial_t u_3 + v_3 \partial_x u_3 +i\varepsilon {u}_1u_2 &=& 0.
\end{eqnarray}\end{mathletters}
Here
\begin{equation} \label{v123}
v_1 = \frac{b_2-b_1}{a_1-a_2},\quad v_2 = \frac{b_3-b_2}{a_2-a_3},\quad v_3 =
\frac{b_3-b_1}{a_1-a_3},
\label{add1}\end{equation}
\begin{equation}\label{varepsilon}
\varepsilon = \frac{a_1b_2-a_2b_1+a_2b_3-a_3b_2+a_3b_1-a_1b_3}
{[(a_1-a_2)(a_2-a_3)(a_1-a_3)]^{1/2}}.
\label{add2}
\end{equation}
The group velocities satisfy the following condition
\begin{equation}
\frac{v_2-v_3}{v_1-v_3} = -\frac{a_1-a_2}{a_2-a_3}<0.
\label{ineqv}\end{equation}
The three-wave system (\ref{3wave}) can be interpreted physically. It
describes the interaction of three wave packets with complex envelopes
$u_1$, $u_2$ and $u_3$ in a medium with quadratic nonlinearity.

It is often desirable to relate the inverse-scattering parameters $a_j$ and
$b_j$ $(j=1, 2, 3)$ to the physical parameters $\varepsilon$ and $v_j\; (j=1,
2, 3)$. This relation can be easily found from (\ref{v123}) and
(\ref{varepsilon}) as
\begin{equation}\label{a123}
a_1-a_2=\frac{\varepsilon^2}{(v_1-v_2)(v_1-v_3)}, \;\;
a_2-a_3=\frac{\varepsilon^2}{(v_1-v_2)(v_3-v_2)}.
\end{equation}
The other parameters $a_1-a_3$ and $b_j \; (j=1, 2, 3)$ can be readily obtained
from equations (\ref{a123}) and (\ref{v123}). Note that the inverse-scattering
parameters  are not uniquely determined. In fact, one of $a_j$ and one of $b_j$
$(j=1, 2, 3)$ are free parameters. It is an invariance in the
inverse-scattering formulation of the 3-wave system and it does not affect the
physical solution in any way.

In general, the Riemann-Hilbert problem (\ref{RH1})-(\ref{RH2}) has multiple
solutions. Different solutions  are related to each other by the rational
matrix functions $\Gamma(k)$  (which also depend on the variables $x$ and $t$)
\cite{NMPZ84,FT87,ZS79a,ZS79b,Kawata}:
\begin{equation}
\widetilde{\Phi}_\pm(k,x,t) = \Phi_\pm(k,x,t)\Gamma(k,x,t).
\label{RH6}\end{equation}
The rational matrix $\Gamma(k)$ must satisfy the canonical
normalization condition: $\Gamma(k)\to I$ for $k\to \infty$ and must
have poles only in $C_-$ (the inverse function  $\Gamma^{-1}(k)$ then
has poles in $C_+$ only). Such a rational matrix $\Gamma(k)$ will be
called the soliton matrix below, since it gives the soliton part of the
solution to the integrable nonlinear PDE.

To specify a unique solution to the Riemann-Hilbert problem the set of the
Riemann-Hilbert data must be given. These data are also called the spectral
data. The full set of the spectral data comprises the matrix $G(k,x,t)$ on the
right-hand side of equation (\ref{RH1}) and the appropriate discrete data
related to the zeros of $\det\Phi_+(k)$ and $\det\Phi_-^{-1}(k)$. We will
confine ourselves to the case of the Riemann-Hilbert problem with zero index,
i.e., when $\det\Phi_+(k)$ and $\det\Phi_-^{-1}(k)$ have equal number of zeros
(counting the multiplicity). For instance, in the case of involution
(\ref{invol}) the Riemann-Hilbert problem has zero index because the zeros
appear in complex conjugate pairs: $\overline{k}_j = k^*_j$. It is known
\cite{BC1,BC2,BC3,BDT,Zhou1,Zhou2} (see also Ref.~\cite{Kawata}) that in the
generic case the spectral data  include  simple (distinct) zeros
$k_1,\ldots,k_n$ of $\det\Phi_+(k)$ and $\overline{k}_1,\ldots,\overline{k}_n$
of $\det\Phi_-^{-1}(k)$, in their holomorphicity domains, and the null vectors
$|v_1\rangle,\ldots,|v_n\rangle$ and $\langle \overline{v}_1|,\ldots,\langle
\overline{v}_n|$ from the respective kernels:
\begin{equation}
\Phi_+(k_j)|v_j\rangle = 0,\quad \langle
\overline{v}_j|\Phi_-^{-1}(\overline{k}_j)=0.
\label{vects}\end{equation}

Using the property (\ref{Traces}) one can verify that the zeros do not depend
on the variables $x$ and $t$. The  $(x,t)$-dependence of the null vectors can
be easily derived by differentiation of (\ref{vects}) and use of the linear
spectral equations (\ref{RH3})-(\ref{RH4}). This dependence reads:
\begin{mathletters}
\label{RH8}
\begin{eqnarray}
|v_j\rangle &=&
\exp{\{-\Lambda(k_j)x-\Omega(k_j)t\}}|v^{(0)}_j\rangle,\\
\langle \overline{v}_j| &=& \langle
\overline{v}^{(0)}_j|\exp{\{\Lambda(\overline{k}_j)x+\Omega(\overline{k}_j)t\}},
\end{eqnarray}
\end{mathletters}
where $|v^{(0)}_j\rangle$ and $\langle \overline{v}^{(0)}_j|$ are some
constant vectors.

The vectors in equation (\ref{RH8}) together with the zeros constitute
the full set of the discrete data necessary to specify the soliton
matrix $\Gamma(k,x,t)$ and,  hence, unique solution to the
Riemann-Hilbert problem (\ref{RH1})-(\ref{RH2}). Indeed, by
constructing the soliton matrix $\Gamma(k)$ such that the following
matrix functions
\begin{equation}
\phi_+(k) = \Phi_+(k)\Gamma^{-1}(k),\quad \phi_-^{-1}(k) =
\Gamma(k)\Phi_-^{-1}(k)
\label{regularphi}\end{equation}
are nondegenerate and holomorphic in the domains $C_+$ and $C_-$,
respectively, we reduce the Riemann-Hilbert problem with zeros to
another one without zeros and hence uniquely solvable (for details see,
for instance, Refs.~\cite{NMPZ84,FT87,AC91,Kawata}). Below by matrix
$\Gamma(k)$ we will imply the matrix from equation (\ref{regularphi})
which reduces the Riemann-Hilbert problem (\ref{RH1})-(\ref{RH2}) to
the one without zeros. The corresponding  solution to the integrable
PDE (\ref{RH5}) is obtained by using the asymptotic expansion of the
matrix $\Phi(k)$ as $k\to\infty$ in the linear equation (\ref{RH3}). In
the $N$-wave interaction model it is given by formula (\ref{UV3wave}).
The pure soliton solutions are obtained by using the rational matrix
$\Phi=\Gamma(k)$.

\section{Soliton matrices for multiple zeros}
\label{multpl}

In this section we consider the soliton solution corresponding to a single
multiple zero of arbitrary order in the case of an arbitrary matrix  dimension
$N$. Such soliton solutions will be referred to as the higher-order solitons.
We will {\em derive} the general formulae for the soliton matrices
corresponding to an elementary higher-order zero (see the definition below)
starting from the usual elementary soliton matrices of the Riemann-Hilbert
problem. Our formulae for the soliton matrices corresponding to an elementary
higher-order zero are similar to the previously proposed {\em ansatz} for the
$2\times2$ Zakharov-Shabat spectral problem  \cite{Nathalie}. However, in our
approach some essential invariance properties and simple evolution formulae for
the vector parameters in the soliton matrices are given, which were not known
before. Thus we simplify the ansatz of Refs.~\cite{OptLett,Nathalie} and put it
on the rigorous footing. Although we work in the case of involution
({\ref{invol}), usual for applications in nonlinear physics, our approach is
valid for the general Riemann-Hilbert problem with zero index. Moreover, we
present our formulae in a form transferable without any changes to that general
case.

Let $\Phi_+(k)$ and $\Phi_-^{-1}(k)$ from (\ref{RH1}) each have but one
zero of order $n$, $k_1$ and $\overline{k}_1$, respectively:
\begin{equation}
\det\Phi_+(k) = (k-k_1)^n\varphi(k),\quad \det\Phi^{-1}_-(k) =
(k-\overline{k}_1)^n\overline{\varphi}(k),
\end{equation}
where  $\det\varphi(k_1)\ne0$ and
$\det\overline{\varphi}(\overline{k}_1)\ne0$. The geometric
multiplicity of $k_1$ ($\overline{k}_1$) is defined as the number of
the null vectors in the kernel of $\Phi_+(k_1)$
($\Phi_-^{-1}(\overline{k}_1)$), see (\ref{vects}). It can be easily
shown that the order of a zero is always greater or equal to its
geometric multiplicity. It is also obvious that the geometric
multiplicity of a zero is less than the matrix dimension. Before we
proceed with the construction of the soliton matrix $\Gamma(k)$
corresponding to the multiple zero of order $n$, two important
properties must be pointed out. It is convenient to formulate them in
the form of two lemmas.

\begin{lem}
\label{lemma1}
Suppose vectors $|v_j\rangle \;(1\le j\le m)$ are in the kernel of
matrix $\Phi_+(k_1)$, i.e.,
\begin{equation}\label{lemma1a}
\Phi_+(k_1)|v_j\rangle = 0,\quad j=1,\ldots,m,
\end{equation}
where $m$ is less or equal to $k_1$'s geometric multiplicity. Define
the new matrix $\widetilde{\Phi}_+(k)\equiv \Phi(k)_+\chi^{-1}(k)$
where
\begin{equation}
\chi(k) = I-\frac{k_1-\overline{k}_1}{k-\overline{k}_1}P,
\label{chidef}
\end{equation}
\begin{equation}
P = \sum_{i, j=1}^{m}|v_i
\rangle(K^{-1})_{ij}\langle\overline{v}_j|,\quad K_{ij} =
\langle\overline{v}_{i}|v_{j}\rangle,
\end{equation}
and vectors $\langle\overline{v}_j| \;(1\le j\le m)$ are arbitrary but
they make matrix $K$ invertible. Then matrix $\widetilde{\Phi}_+(k)$ is
also holomorphic in the upper half plane. In addition, if a new vector
$|w\rangle$ is in the kernel of $\widetilde{\Phi}_+(k_1)$ and is
orthogonal to $\langle \overline{v}_j| \; (1\le j\le m)$, i.e.,
\begin{equation}
\widetilde{\Phi}_+(k_1)|w\rangle = 0,\quad \langle \overline{v}_j | w
\rangle=0, \quad j=1,\ldots,m,
\end{equation}
then
\begin{equation}
\Phi_+(k_1)| w\rangle =0,
\end{equation}
i.e., $|w\rangle$ is also in the kernel of $\Phi_+(k_1)$.  Furthermore,
$|w\rangle$ is linearly independent of $|v_j\rangle \; (1\le j\le m)$. Similar
results exist for matrix $\Phi_-^{-1}(k)$ where the multiplication is one the
left.
\end{lem}

\noindent Remark: it is easy to see that in order for $K$ to be
invertible, it is necessary that vectors $\langle\overline{v}_j|
\;(1\le j\le m)$ be linearly independent. But this condition is not
sufficient. However, if $\langle\overline{v}_j|=|v_j\rangle^\dag \;
(j=1, \dots, m)$, then it can be shown that $K$ is invertible.

\vspace{0.3cm}
\noindent{\it Proof}.
The matrix $P$ is clearly a projector matrix, thus
\begin{equation} \label{chiinv}
\chi^{-1}(k) = I+\frac{k_1-\overline{k}_1}{k-k_1}P.
\end{equation}
Then, expanding the holomorphic function $\Phi^+(k)$ into the Taylor
series and recalling equation (\ref{lemma1a}), we see that
\begin{eqnarray}
\Phi^+(k)\chi^{-1}(k)=&\left\{\Phi^+(k_1)+(k-k_1)\frac{d\Phi^+(k_1)}{dk}
+(k-k_1)^2\frac{d^2\Phi^+(k_1)}{2!dk^2}+\dots \right\}
(1+\frac{k_1-\overline{k}_1}{k-k_1}P) \nonumber \\
=& \Phi^+(k_1)+(k_1-\overline{k}_1)\frac{d\Phi^+(k_1)}{dk}P
+(k_1-\overline{k}_1)(k-k_1)\frac{d^2\Phi^+(k_1)}{2!dk^2}P+\dots,
\end{eqnarray}
which is clearly holomorphic.

Next, if
\[
\widetilde{\Phi}_+(k_1)|w\rangle = 0,
\]
recalling the definition of $\widetilde{\Phi}_+(k)$ and expanding
$\Phi^+(k)$ into the Taylor series, we get
\begin{equation}
\Phi^+(k_1)|w\rangle
+(k_1-\overline{k}_1)\frac{d\Phi^+(k_1)}{dk}P|w\rangle=0.
\end{equation}
Since by assumption,
\[\langle \overline{v}_j | w \rangle=0, \quad j=1,\ldots,m,\]
thus,
\[P |w\rangle=0, \]
consequently,
\begin{equation}
\Phi^+(k_1)|w\rangle=0.
\end{equation}
 Lastly, $|w\rangle$ is linearly independent of $|v_j\rangle \; (j=1, \dots,
m)$ because the matrix $K$ is invertible.  Q.E.D.

{\bf Corollary 1} \hspace{0.1cm} Suppose the kernel of $\Phi_+(k_1)$ is spanned
by vectors $|v_j\rangle\; (1\le j\le m)$ where $m$ is the geometric
multiplicity of zero $k_1$. Define matrices $\chi(k)$, $\tilde{\Phi}_+(k)$ and
projector $P$ as in Lemma \ref{lemma1}. Then, there exists no vector in the
kernel of $\tilde{\Phi}_+(k_1)$ which is simultaneously orthogonal to $\langle
\overline{v}_j|\; (1\le j\le m)$.

\begin{lem}
\label{lemma2}
Suppose that $\Phi_+(k_1)$ has $r$ independent vectors in the kernel:
\begin{equation}
\Phi_+(k_1)|v_j\rangle = 0,\quad j=1,\ldots,r,
\label{S1}\end{equation}
i.e., ${\rm rank\,}\Phi_+(k_1)=N-r$. Then the following matrix function
$\widetilde{\Phi}_+(k)\equiv \Phi_+(k)\chi^{-1}(k)$, where matrix
$\chi(k)$ is as defined in equation (\ref{chidef}) but with
\begin{equation}
P = \sum_{i, j=1}^{r}|v_i
\rangle(K^{-1})_{ij}\langle\overline{v}_{j}|,\quad K_{ij} =
\langle\overline{v}_{i}|v_{j}\rangle,
\end{equation}
has at most $r$ vectors in the kernel at $k=k_1$, i.e., ${\rm
rank\,}\widetilde{\Phi}_+(k_1) \ge N-r$. Here vectors $\langle
\overline{v}_j| \; (1\le j\le r)$ are arbitrary but they make matrix
$K$ invertible.
\end{lem}

\noindent{\it Proof}. This lemma is easy to prove by contradiction.
Suppose that there are at least $r+1$ independent vectors
$|u_1\rangle$, ..., $|u_{r+1}\rangle$ in the kernel of
$\widetilde{\Phi}_+(k_1)$ defined above. Then one can find a non-zero
vector $|X\rangle$ in the kernel of $\widetilde{\Phi}_+(k_1)$ such that
\begin{equation} \label{vX}
\langle\overline{v}_j|X\rangle = 0,\quad j=1,\ldots,r.
\label{3.13}\end{equation} Indeed, substitution of the expansion
\[
|X\rangle = \sum_{j=1}^{r+1}C_j|u_j\rangle
\]
into Eq. (\ref{vX}) leads to an underdetermined, hence, solvable system
of equations
\[
\sum_{j=1}^{r+1}\langle\overline{v}_i|u_j\rangle C_j = 0,\quad
i=1,\dots,r
\]
which have non-zero solutions. But then, according to the second part
of Lemma~1, $|X\rangle$ is also in the kernel of $\Phi_+(k_1)$, thus
\begin{equation} \label{X}
|X\rangle = \sum_{j=1}^rC_j|v_j\rangle.
\end{equation}
Substituting Eq. (\ref{X}) into (\ref{3.13}) and recalling that the
matrix $K$ is invertible, we find that $C_j=0, \; j=1, \dots, r$, hence
$X=0$. Thus we have arrived at a contradiction. Q.E.D. (Note that a
similar lemma is valid for $\Phi_-^{-1}(k)$ at $k=\overline{k}_1$ with
the multiplication on the left.)

To clarify the implications of Lemma \ref{lemma2} for the soliton
matrix $\Gamma(k)$ of the higher-order zeros, $k=k_1$ of
$\det\Phi_+(k)$ and $k=\overline{k}_1$ of $\det\Phi_-^{-1}(k)$, let us
examine the way such matrix is constructed. Starting from the solution
$\Phi_\pm(k)$ to the Riemann-Hilbert problem (\ref{RH1})-(\ref{RH2}),
one looks for the independent null vectors for the matrices
$\Phi_+(k_1)$ and $\Phi_-^{-1}(\overline{k}_1)$:
\begin{equation}
\Phi_+(k_1)|v_{i1}\rangle = 0,\quad
\langle\overline{v}_{i1}|\Phi_{-}^{-1}(\overline{k}_1) = 0,\quad i =
1,\ldots,s_1,
\label{S3}\end{equation}
where $s_1$ is the smaller of $k_1$ and $\overline{k}_1$'s geometric
multiplicities. Here we allow the two geometric multiplicities to be
different in general, but they are always the same in the case of
involution (\ref{invol}). Next, one constructs the elementary matrix
\begin{equation}
\chi_1(k) = I -\frac{k_1-\overline{k}_1}{k-\overline{k}_1}P_1,\quad
\label{S4}\end{equation}
where
\begin{equation}
P_1 = \sum_{i, j}^{s_1}|v_{i1}
\rangle(K^{-1})_{ij}\langle\overline{v}_{j 1}|,\quad K_{ij} =
\langle\overline{v}_{i1}|v_{j 1}\rangle.
\label{S5}\end{equation}
It can be shown that
$\mbox{det}\chi_1=\left(\frac{k-k_1}{k-\overline{k}_1}\right)^{s_1}$.
If $s_1<n$, where $n$ is the order of the two zeros, then one considers
the matrix functions $\widetilde{\Phi}_+(k) = \Phi_+(k)\chi_1^{-1}(k)$
and $\widetilde{\Phi}_-^{-1}(k)=\chi_1(k)\Phi_-^{-1}(k)$. From Lemma
\ref{lemma1}, we know that matrices $\widetilde{\Phi}_+(k)$ and
$\widetilde{\Phi}_-^{-1}(k)$ are also holomorphic in the respective
half planes of the complex plane. In addition, $k_1$ ($\overline{k}_1$)
is still a zero of $\mbox{det}\widetilde{\Phi}_+(k)$
($\mbox{det}\widetilde{\Phi}_-^{-1}(k)$). Repeating the above steps one
gets the elementary matrices $\chi_1(k)$, \ldots, $\chi_r(k)$ such that
$s_1+s_2+\ldots +s_r = n$. Therefore,
\begin{equation}
\Gamma(k) = \chi_{r}(k)\cdot\ldots\cdot\chi_2(k)\chi_1(k),
\label{S6}\end{equation}
where matrices $\chi_l(k)$ and projectors $P_l$ are as defined in
equations (\ref{S4}) and (\ref{S5}) but the independent vectors
$|v_{il}\rangle$ and $\langle\overline{v}_{il}|$ ($i=1, \dots, s_l$)
are from the kernels of
$(\Phi_+\chi^{-1}_1\cdot\ldots\cdot\chi^{-1}_{l-1})(k_1)$ and
$(\chi_{l-1}\cdot\dots\cdot\chi_1\Phi_-^{-1})(\overline{k}_1)$
respectively.

Lemma \ref{lemma2} indicates that in fact the sequence of ranks of the
projectors $P_l$ in the matrix $\Gamma(k)$ given by equation
(\ref{S6}), i.e. built in the described way, is non-increasing:
\begin{equation}\label{rankPr}
\text{rank\,}P_r\le\text{rank\,}P_{r-1}\le\ldots\le\text{rank\,}P_1.
\label{S7}\end{equation}
This result allows one to classify possible occurrences of a
higher-order zero of the Riemann-Hilbert problem for arbitrary matrix
dimension $N$.   In general, for zeros of the same order $n$, different
sequences of ranks in formula (\ref{S7}) give different classes of the
higher-order soliton solutions.  In the present paper we consider in
detail only the higher-order zeros when the sequence of ranks
(\ref{S7}) is the simplest possible: $\text{rank\,}P_l = 1$,
$l=1,\ldots,n$. We introduce the following definition.
\medskip

\noindent
{\bf Definition 1.} {\sl In the soliton matrix (\ref{S6}) corresponding to a
higher-order zero $k_1$  of a Riemann-Hilbert problem, if the ranks of all
projectors $P_l (1\le l\le n)$ are 1, then we call this zero an elementary
higher-order zero.}

Remark 1: We observe from equation (\ref{rankPr}) that  a higher-order zero  (of
arbitrary algebraic multiplicity) is elementary if and only if $\text{rank}P_1
= 1$, i.e., the geometric multiplicity of the zero is 1.

Remark 2: If the matrix dimension $N=2$ (as for the nonlinear Schr\"odinger
equation), then all higher-order zeros are elementary since $\text{rank}P_1$ is
always equal to 1.

Below we derive the soliton matrix $\Gamma(k)$ and its inverse for an
elementary higher-order zero. The results are presented in the following
lemma.

\begin{lem}
\label{lemma3}
Consider a pair of elementary higher-order zeros of order $n$: $k=k_1$ in
$C_+$ and $k=\overline{k}_1$ in $C_-$. Then the corresponding soliton
matrix $\Gamma(k)$ and its inverse can be cast in the following form:
\begin{mathletters}
\label{S8}
\begin{equation}
\Gamma(k) = I +
\sum_{l=1}^n\sum_{j=1}^l\frac{|\overline{q}_j\rangle\langle
\overline{p}_{l+1-j}|} {(k-\overline{k}_1)^{n+1-l}} =
I+(|\overline{q}_n\rangle,\ldots,|\overline{q}_1\rangle)
\overline{D}(k)\left(\begin{array}{c} \langle \overline{p}_1| \\ \vdots
\\ \langle \overline{p}_n| \end{array}\right),\label{S8a}\end{equation}
\begin{equation}
\Gamma^{-1}(k) = I +
\sum_{l=1}^n\sum_{j=1}^l\frac{|p_{l+1-j}\rangle\langle
q_j|}{(k-{k}_1)^{n+1-l}} = I+(|p_1\rangle,\ldots,|p_n\rangle)D(k)
\left(\begin{array}{c} \langle{q}_n| \\ \vdots \\ \langle {q}_1|
\end{array}\right),\label{S8b}\end{equation}
\end{mathletters}
where the matrices $D(k)$ and $\overline{D}(k)$ are defined as
\begin{equation}
\overline{D}(k) = \left(\begin{array}{cccc}
\frac{1}{(k-\overline{k}_1)}&0&\ldots&0\\
\frac{1}{(k-\overline{k}_1)^{2}}&\frac{1}{(k-\overline{k}_1)}&\ddots&\vdots\\
\vdots&\ddots&\ddots&0\\
\frac{1}{(k-\overline{k}_1)^n}& \ldots &
\frac{1}{(k-\overline{k}_1)^{2}}
&\frac{1}{(k-\overline{k}_1)}\end{array}\right),\; D(k) =
\left(\begin{array}{cccc} \frac{1}{(k-k_1)}&\frac{1}{(k-k_1)^2}&\ldots&
\frac{1}{(k-k_1)^n}\\
0&\ddots&\ddots&\vdots \\
\vdots&\ddots&\frac{1}{(k-k_1)}&\frac{1}{(k-k_1)^2}\\
0&\ldots&0&\frac{1}{(k-k_1)}\end{array}\right),
\label{Ds}\end{equation}
and vectors $| p_j\rangle, \langle\overline{p}_j|, \langle q_j|,
|\overline{q}_j\rangle \; (j=1, \dots, n)$ are independent of $k$.
\end{lem}

 Remark: In \cite{Nathalie}, the ansatz of the form (\ref{S8}) was proposed
for higher-order solitons in the nonlinear Schr\"odinger equation. The lemma
above, together with Remark 2 below Definition 1, shows that their ansatz is in
fact the most general soliton matrix for the nonlinear Schr\"odinger equation.
If $N>2$, their ansatz then is just the soliton matrix for {\em elementary}
higher-order zeros.

\noindent{\it Proof}. The representation (\ref{S8}) can be proved by
induction. Consider, for instance, formula (\ref{S8a}). Obviously, this
formula is valid for $n=1$. In this case, $\Gamma(k)$ reduces to an
elementary matrix $\chi(k)$. Now, suppose that formula (\ref{S8a}) is
valid for $n=m$. Then we need to show that it is valid for $n=m+1$ as
well. Indeed, denote the soliton matrices for $n=m$ and $n=m+1$ by
$\Gamma(k)$ and $\widetilde{\Gamma}(k)$ respectively. Then taking into
account expression (\ref{S6}) and recalling our assumption of the
elementary higher-order zero, we have
\begin{equation}
\widetilde{\Gamma}(k) = \chi_{m+1}(k)\Gamma(k) =
\left(I+\frac{|v_{m+1}\rangle
\langle\overline{v}_{m+1}|}{k-\overline{k}_1}\right)\left( I +
\sum_{l=1}^m\sum_{j=1}^l\frac{|\overline{q}_j\rangle\langle
\overline{p}_{l+1-j}|} {(k-\overline{k}_1)^{m+1-l}} \right).
\label{S9}\end{equation}
Here, for simplicity of the formulae below, we have normalized the
vectors $|v_{m+1}\rangle$ and $\langle\overline{v}_{m+1}|$ such that
$\langle\overline{v}_{m+1}|v_{m+1}\rangle = \overline{k}_1-k_1$. Let us
now multiply the two terms in the right-hand side of equation
(\ref{S9}) and compute the coefficients at the poles:
\begin{equation}
\widetilde{\Gamma}(k)= I +
\frac{\widetilde{A}_1}{k-\overline{k}_1}+\frac{\widetilde{A}_2}{(k-\overline{k}_1)^2}
+ \ldots+\frac{\widetilde{A}_{m+1}}{(k-\overline{k}_1)^{m+1}},
\label{S10}\end{equation}
where
\begin{eqnarray*}
\widetilde{A}_{m+1} &=& |v_{m+1}\rangle\langle\overline{v}_{m+1}|A_{m}
=
|v_{m+1}\rangle\langle\overline{v}_{m+1}|\overline{q}_1\rangle\langle\overline{p}_1|,
\\
\;\widetilde{A}_{m}\; &=&
|v_{m+1}\rangle\langle\overline{v}_{m+1}|A_{m-1}+ A_m  \\
&=& |v_{m+1}\rangle\langle\overline{v}_{m+1}| \Bigl(
|\overline{q}_2\rangle\langle\overline{p}_1|+|\overline{q}_1\rangle\langle\overline{p}_2|
\Bigr)
+ |\overline{q}_1\rangle\langle\overline{p}_1| \\
&=&
\left(|v_{m+1}\rangle\langle\overline{v}_{m+1}|\overline{q}_2\rangle +
|\overline{q}_1\rangle\right)\langle\overline{p}_1| +
|v_{m+1}\rangle\langle\overline{v}_{m+1}|
\overline{q}_1\rangle\langle\overline{p}_2|,
\\
\widetilde{A}_{m-1} &=&
|v_{m+1}\rangle\langle\overline{v}_{m+1}|A_{m-2}+ A_{m-1} \\
&=&  |v_{m+1}\rangle\langle\overline{v}_{m+1}| \Bigl(
|\overline{q}_3\rangle\langle\overline{p}_1| +
|\overline{q}_2\rangle\langle\overline{p}_2| +
|\overline{q}_1\rangle\langle\overline{p}_3| \Bigr) +
|\overline{q}_2\rangle\langle\overline{p}_1| +
|\overline{q}_1\rangle\langle\overline{p}_2|  \\
&=
&\left(|v_{m+1}\rangle\langle\overline{v}_{m+1}|\overline{q}_3\rangle +
|\overline{q}_2\rangle\right)\langle\overline{p}_1| +
\left(|v_{m+1}\rangle\langle\overline{v}_{m+1}|\overline{q}_2\rangle
+ |\overline{q}_1\rangle\right)\langle\overline{p}_2|  \\
& & + |v_{m+1}\rangle\langle\overline{v}_{m+1}|
\overline{q}_1\rangle\langle\overline{p}_3|,\\
& &\ldots, \\
\;\widetilde{A}_{1}\; &=& |v_{m+1}\rangle\langle\overline{v}_{m+1}| +
\sum_{j=1}^m |\overline{q}_{m+1-j}\rangle\langle\overline{p}_j|.
\end{eqnarray*}
Define new vectors:
\[
|\widetilde{\overline{q}}_1\rangle  =
|v_{m+1}\rangle\langle\overline{v}_{m+1}| {\overline{q}}_1\rangle,\quad
|\widetilde{\overline{q}}_j\rangle =
|v_{m+1}\rangle\langle\overline{v}_{m+1}| {\overline{q}}_j\rangle +
|{\overline{q}}_{j-1}\rangle,\quad j=2,\ldots,m, \quad
\]
\begin{equation}
 |\widetilde{\overline{q}}_{m+1}\rangle = |\overline{q}_m\rangle,\quad
\langle\overline{p}_{m+1}| = \frac{\langle \overline{v}_{m+1}|
-\sum_{j=1}^{m-1}\langle\overline{v}_{m+1}|\overline{q}_{j+1}
\rangle\langle\overline{p}_{m-j+1}|}{\langle\overline{v}_{m+1}|\overline{q}_1\rangle}.
\label{S11}\end{equation}
Then matrices $\widetilde{A}_{1},\ldots,\widetilde{A}_{m+1}$ take the
following representation:
\begin{equation}
\widetilde{A}_{m+2-l} = \sum_{j=1}^{l}|\widetilde{\overline{q}}_{l+1-j}
\rangle\langle\overline{p}_j|,\quad l=1,\ldots,m+1.
\label{S12}
\end{equation}
Thus formula (\ref{S8a}) is valid for $\widetilde{\Gamma}(k)$ as well.
It is noted that we must also show that the denominator in formula
(\ref{S11}) is nonzero. This is easy to show, as $\langle
\overline{v}_{m+1}|\overline{q}_1 \rangle$ is actually (up to a factor
$(\overline{k}_1-k_1)^m$) a product of inner products $\langle
\overline{v}_{j+1}|v_j\rangle \; (1\le j\le m)$, where $|v_j\rangle$
and $\langle\overline{v}_j|$ are the projector vectors in matrix
$\chi_j$:
\begin{equation}
\chi_j(k)=I-\frac{k_1-\overline{k}_1}{k-\overline{k}_1}|v_j\rangle
\langle \overline{v}_j|
\end{equation}
[see equations (\ref{S6}) and (\ref{S9})]. If $\langle
\overline{v}_{j+1}|v_j\rangle=0$ for some $j$, then Lemma \ref{lemma1}
indicates that the projector $P_j$ in matrix $\chi_j$ [see (\ref{S4})
to (\ref{S6})] would have rank higher than 1, which contradicts our
assumption of elementary higher-order zeros. Thus $\langle
\overline{v}_{j+1} | v_j \rangle\ne 0$ for all $j$, consequently,
$\langle\overline{v}_{m+1}|\overline{q}_1\rangle \ne 0$. Expression for
$\Gamma^{-1}(k)$ (\ref{S8b}) can be proved in the same way. Q.E.D.

In the expressions for $\Gamma(k)$ (\ref{S8a}) and $\Gamma^{-1}(k)$
(\ref{S8b}) there are twice as many vectors as in the elementary
matrices (\ref{S4}) and (\ref{S6}). As the result, only half of the
vector parameters, namely $|p_1\rangle,\ldots,|p_n\rangle$ and
$\langle\overline{p}_1|,\ldots,\langle\overline{p}_n|$, are
independent. To derive the formulae for the rest of the vector
parameters in (\ref{S8}) we can use the identity
$\Gamma(k)\Gamma^{-1}(k) = I$. The poles of $\Gamma(k)\Gamma^{-1}(k)$
at $k=k_1$, starting from the highest order pole, give:
\begin{eqnarray*}
& &\Gamma(k_1)|p_1\rangle\langle q_1| = 0,\\
& & \Gamma(k_1)\left(|p_2\rangle\langle q_1| + |p_1\rangle\langle
q_2|\right)
+\frac{1}{1!}\frac{\text{d}\Gamma(k_1)}{\text{d}k}|p_1\rangle\langle
q_1| = 0,\\
& & \Gamma(k_1)\left(|p_3\rangle\langle q_1| + |p_2\rangle\langle q_2|
+ |p_1\rangle\langle q_3|\right) +
\frac{1}{1!}\frac{\text{d}\Gamma(k_1)}{\text{d}k}
\left(|p_2\rangle\langle q_1| + |p_1\rangle\langle q_2|\right)
 + \frac{1}{2!}\frac{\text{d}^2\Gamma(k_1)}{\text{d}k^2}
|p_1\rangle\langle q_1| = 0,\\
& &\qquad\ldots \,.\end{eqnarray*} Hence, we obtain:
\begin{mathletters}
\label{S13}
\begin{eqnarray}
& &\Gamma(k_1)|p_1\rangle = 0,\\
& &\Gamma(k_1)|p_2\rangle
+\frac{1}{1!}\frac{\text{d}\Gamma(k_1)}{\text{d}k}|p_1\rangle = 0,\\
& &\Gamma(k_1)|p_3\rangle +
\frac{1}{1!}\frac{\text{d}\Gamma(k_1)}{\text{d}k} |p_2\rangle +
\frac{1}{2!}\frac{\text{d}^2\Gamma(k_1)}{\text{d}k^2}
|p_1\rangle = 0, \\
& &\qquad \ldots, \nonumber\\
& & \Gamma(k_1)|p_n\rangle +
\frac{1}{1!}\frac{\text{d}\Gamma(k_1)}{\text{d}k}|p_{n-1}\rangle +
\ldots
+\frac{1}{(n-1)!}\frac{\text{d}^{n-1}\Gamma(k_1)}{\text{d}k^{n-1}}|p_1\rangle
= 0.
\end{eqnarray}\end{mathletters}
Equations (\ref{S13}) can be written in a compact form for the following matrix
${\bf\Gamma}(k)$:
\begin{equation}
{\bf\Gamma}(k_1)\left(\begin{array}{c} |p_1\rangle \\ \vdots \\
|p_n\rangle
\end{array}\right) = 0, \qquad
{\bf\Gamma}(k) \equiv \left(\begin{array}{cccc}
\Gamma&0&\ldots&\quad 0\\
\frac{1}{1!}\frac{\text{d}}{\text{d}k}\Gamma&\Gamma&\ddots&\quad\vdots\\
\vdots&\ddots&\ddots&\quad 0\\
\frac{1}{(n-1)!}\frac{\text{d}^{n-1}}{\text{d}k^{n-1}}\Gamma& \ldots &
\frac{1}{1!}\frac{\text{d}}{\text{d}k}\Gamma &\quad\Gamma
\end{array}\quad\right).
\label{S14}\end{equation}
Note that, as a block matrix, ${\bf\Gamma}(k)$ has (lower-triangular) Toeplitz
form, i.e. along each diagonal it has the same (matrix) element.

In much the same way, by considering the poles at $k=\overline{k}_1$ in
$\Gamma(k)\Gamma^{-1}(k)$, one derives the following formula
\begin{equation}
\left(\langle\overline{p}_1|,\ldots,\langle\overline{p}_n|\right)
\overline{\bf\Gamma}(\overline{k}_1) = 0,\qquad \overline{\bf\Gamma}(k)
= \left(\begin{array}{cccc} \Gamma^{-1}\quad
&\frac{1}{1!}\frac{\text{d}}{\text{d}k}\Gamma^{-1}\quad &\ldots\quad &
\frac{1}{(n-1)!}\frac{\text{d}^{n-1}}{\text{d}k^{n-1}}\Gamma^{-1}\\
0\quad &\Gamma^{-1}\quad &\ddots\quad & \vdots  \\
\vdots\quad &\ddots\quad &\ddots\quad &
\frac{1}{1!}\frac{\text{d}}{\text{d}k}\Gamma^{-1} \\
0\quad &\ldots \quad &0 \quad &\Gamma^{-1}
\end{array}\right).
\label{S15}\end{equation}

Equations (\ref{S14}) and (\ref{S15}) allow us to find the expressions
for the dependent vector parameters. For convenience of the
presentation, let us introduce the following $k$-dependent vectors:
\begin{equation}
\langle\overline{Z}_j(k)| =
\sum_{l=j}^n\frac{\langle\overline{p}_{l+1-j}|}{(k-\overline{k}_1)^{n+1-l}}\,,
\qquad |{Z}_j(k)\rangle =
\sum_{l=j}^n\frac{|p_{l+1-j}\rangle}{(k-k_1)^{n+1-l}}\,.
\label{S16}\end{equation}
Then, by reordering the summation in (\ref{S8}) we get
\begin{mathletters}
\label{S17}
\begin{equation}
\Gamma(k) =
I+(|\overline{q}_n\rangle,\ldots,|\overline{q}_1\rangle)\left(\begin{array}{c}
\langle \overline{Z}_n(k)|\\ \vdots\\ \langle
\overline{Z}_1(k)|\end{array}\right),
\label{S17a}\end{equation}
\begin{equation}
\Gamma^{-1}(k) =
I+(|{Z}_n(k)\rangle,\ldots,|{Z}_1(k)\rangle)\left(\begin{array}{c}
\langle{q}_n| \\ \vdots \\ \langle {q}_1|
\end{array}\right).
\label{S17b}\end{equation}
\end{mathletters}
Let us now substitute the expression (\ref{S17a}) into equation
(\ref{S14}) and solve for
$|\overline{q}_1\rangle,\ldots,|\overline{q}_n\rangle$. We have
\begin{eqnarray*}
& &|p_1\rangle + (|\overline{q}_n\rangle,\ldots,|\overline{q}_1\rangle)
\left(\begin{array}{c} \langle \overline{Z}_n(k_1)|p_1\rangle\\
\vdots\\ \langle \overline{Z}_1(k_1)|p_1\rangle
\end{array}\right) = 0,\\
& &|p_2\rangle + (|\overline{q}_n\rangle,\ldots,|\overline{q}_1\rangle)
\left(\begin{array}{c} \langle \overline{Z}_n(k_1)|p_2\rangle +
\frac{1}{1!}\frac{\text{d}}{\text{d}k}\langle
\overline{Z}_n(k_1)|p_1\rangle\\
 \vdots\\ \langle
\overline{Z}_1(k_1)|p_2\rangle +
\frac{1}{1!}\frac{\text{d}}{\text{d}k}\langle
\overline{Z}_1(k_1)|p_1\rangle
\end{array}\right) = 0,\\
& &|p_3\rangle + (|\overline{q}_n\rangle,\ldots,|\overline{q}_1\rangle)
\left(\begin{array}{c} \langle \overline{Z}_n(k_1)|p_3\rangle +
\frac{1}{1!}\frac{\text{d}}{\text{d}k}
\langle\overline{Z}_n(k_1)|p_2\rangle +
\frac{1}{2!}\frac{\text{d}^2}{\text{d}k^2}\langle
\overline{Z}_n(k_1)|p_1\rangle
\\ \vdots\\ \langle \overline{Z}_1(k_1)|p_3\rangle +
\frac{1}{1!}\frac{\text{d}}{\text{d}k}\langle
\overline{Z}_1(k_1)|p_2\rangle +
\frac{1}{2!}\frac{\text{d}^2}{\text{d}k^2}\langle
\overline{Z}_1(k_1)|p_1\rangle
\end{array}\right) = 0,\\
& &\quad \ldots\,.\end{eqnarray*} Hence
\begin{equation}
(|\overline{q}_n\rangle,\ldots,|\overline{q}_1\rangle) =
-(|p_1\rangle,\ldots,|p_n\rangle)\overline{\cal K}^{-1},
\label{S18}\end{equation}
where
\begin{equation}
\overline{\cal K} = \left(\begin{array}{cccc} \langle
\overline{Z}_n(k_1)|p_1\rangle&\langle \overline{Z}_n(k_1)|p_2\rangle +
\frac{1}{1!}\frac{\text{d}}{\text{d}k} \langle
\overline{Z}_n(k_1)|p_1\rangle&\ldots&\sum\limits_{l=1}^n\frac{1}{(n-l)!}
\frac{\text{d}^{n-l}}{\text{d}k^{n-l}}\langle
\overline{Z}_n(k_1)|p_l\rangle\\
\vdots&\vdots& &\vdots \\
\langle \overline{Z}_1(k_1)|p_1\rangle&\langle
\overline{Z}_1(k_1)|p_2\rangle + \frac{1}{1!}\frac{\text{d}}{\text{d}k}
\langle
\overline{Z}_1(k_1)|p_1\rangle&\ldots&\sum\limits_{l=1}^n\frac{1}{(n-l)!}
\frac{\text{d}^{n-l}}{\text{d}k^{n-l}}\langle
\overline{Z}_1(k_1)|p_l\rangle  \end{array}\right).
\label{S19}\end{equation}
Similarly, we get
\begin{equation}
\left(\begin{array}{c} \langle{q}_n| \\ \vdots \\ \langle {q}_1|
\end{array}\right) =
- {\cal K}^{-1}\left(\begin{array}{c} \langle \overline{p}_1| \\ \vdots
\\
\langle \overline{p}_n| \end{array}\right),
\label{S20}\end{equation}
where
\begin{equation}
{\cal K} = \left(\begin{array}{ccc}
\langle\overline{p}_1|{Z}_n(\overline{k}_1)\rangle&\ldots&
\langle\overline{p}_1|{Z}_1(\overline{k}_1)\rangle\\
\langle\overline{p}_2|{Z}_n(\overline{k}_1)\rangle +
\frac{1}{1!}\frac{\text{d}}{\text{d}k}\langle\overline{p}_1|{Z}_n(\overline{k}_1)\rangle
&\ldots&\langle\overline{p}_2|{Z}_1(\overline{k}_1)\rangle +
\frac{1}{1!}\frac{\text{d}}{\text{d}k}\langle\overline{p}_1|{Z}_1(\overline{k}_1)\rangle
\\
\vdots&
&\vdots\\\sum\limits_{l=1}^n\frac{1}{(n-l)!}\frac{\text{d}^{n-l}}{\text{d}k^{n-l}}
\langle\overline{p}_l|{Z}_n(\overline{k}_1)\rangle
&\ldots&\sum\limits_{l=1}^n\frac{1}{(n-l)!}\frac{\text{d}^{n-l}}{\text{d}k^{n-l}}
\langle\overline{p}_l|{Z}_1(\overline{k}_1)\rangle
\end{array}\right).
\label{S21}\end{equation}

In terms of the independent vector parameters, the soliton matrices
(\ref{S8a}) and (\ref{S8b}) can be rewritten as
\begin{equation}
\Gamma(k) = I-(|p_1\rangle,\ldots,|p_n\rangle)\overline{\cal K}{}^{-1}
\overline{D}(k) \left(\begin{array}{c} \langle \overline{p}_1| \\
\vdots
\\ \langle \overline{p}_n| \end{array}\right),
\label{newform}\end{equation}
\begin{equation}
\Gamma^{-1}(k) = I-(|p_1\rangle,\ldots,|p_n\rangle)D(k){\cal K}^{-1}
\left(\begin{array}{c} \langle \overline{p}_1| \\ \vdots
\\ \langle \overline{p}_n| \end{array}\right),
\label{newform1}\end{equation} where matrices ${\cal K}$ and
$\overline{\cal K}$ are given in equations (\ref{S21}) and (\ref{S19}).

The soliton matrices given by (\ref{newform}) and (\ref{newform1})
possess invariance properties. The invariance is the transformation of
the independent vector parameters which preserves the form of the
soliton matrices and equations defining the vector parameters, i.e.
equations (\ref{S14})-(\ref{S15}).  Let us first consider
transformations of vectors $|p_j\rangle \; (j=1, \dots, n)$. Suppose
these vectors are transformed as
\begin{equation}
(|p_1\rangle,\ldots,|p_n\rangle)
=(|\widetilde{p}_1\rangle,\ldots,|\widetilde{p}_n\rangle)B,
\label{B}
\end{equation}
where $B$ is a $k$-independent matrix which, in general, depends on $(x,t)$.
Here the vectors $\langle\overline{p}_j|\; (j=1, \dots, n)$ remain intact.
Simple calculations show that the new vectors
$|\widetilde{p}_1\rangle,\ldots,|\widetilde{p}_n\rangle$ satisfy equation
(\ref{S14}) if and only if the matrix $B$ has upper-triangular Toeplitz form,
\begin{equation}
B = \left(\begin{array}{cccccc}
b_1 &\quad b_2 &\quad\ldots& \quad\ldots &\quad b_n \\
0 &\quad b_1 & \quad b_2 &\quad \ldots &\quad \vdots \\
\vdots&\quad 0 &\quad\ddots\;&\quad\ddots&\quad\vdots\\
\vdots&\quad\vdots&\quad\ddots&\quad\ddots&\quad b_2& \\
0&\quad\ldots&\quad \ldots&\quad 0 &\quad b_1 \end{array}\right).
\label{Bform}
\end{equation}
Further, we note that under the transformation (\ref{B})-(\ref{Bform})
the matrix $\overline{\cal K}$ transforms as
\begin{equation}
\overline{\cal K}=\widetilde{\overline{\cal K}}B,
\label{barKtilde}\end{equation}
where matrix $\widetilde{\overline{\cal K}}$ is as given by equation
(\ref{S19}) but with vectors $|p_j\rangle$ replaced by the new vectors
$|\widetilde{p}_j\rangle$. From formulae (\ref{B}) and (\ref{barKtilde}) it is
seen that the form (\ref{newform}) of matrix $\Gamma(k)$ is preserved. We still
need to show that for matrix $B$ of the form (\ref{Bform}), the transformation
(\ref{B}) also preserves the form (\ref{newform1}) of matrix $\Gamma^{-1}(k)$.
Notice that matrix $D(k)$ also has upper-triangular Toeplitz form, thus $D(k)$
and $B$ are commutable. Utilizing this property, we can easily show that under
the transformation (\ref{B}), matrix ${\cal K}$ transforms as
\begin{equation}
{\cal K}=\widetilde{\cal K}B,
\end{equation}
where $\widetilde{\cal K}$ is given by equation (\ref{S21}) but with
$|p_j\rangle$ replaced by $|\widetilde{p}_j\rangle$. Thus the form of
matrix $\Gamma^{-1}(k)$ is also preserved. In short, soliton matrices
(\ref{newform}) and (\ref{newform1}) are invariant under the
transformation (\ref{B}) with matrix $B$ given by (\ref{Bform}).

Similarly, we can show that soliton matrices (\ref{newform}) and
(\ref{newform1}) are also invariant under the transformation
\begin{equation}
\left(\begin{array}{c} \langle \overline{p}_1| \\ \vdots \\ \langle
\overline{p}_n| \end{array} \right) =
\overline{B}\left(\begin{array}{c} \langle \widetilde{\overline{p}}_1|
\\ \vdots \\ \langle \widetilde{\overline{p}}_n|
\end{array}\right),
\end{equation}
where the $k$-independent matrix $\overline{B}$ (which, in general, depends on
$(x,t)$) has lower-triangular Toeplitz form,
\begin{equation} \label{Bbarform}
\overline{B} = \left(\begin{array}{ccccc}
\overline{b}_1 \quad &0 \quad & \ldots   \quad &\ldots \quad & 0\\
\overline{b}_2  \quad &\overline{b}_1  \quad &0  \quad&\ddots\quad
&\vdots\\
\vdots  \quad & \overline{b}_2  \quad &\ddots  \quad &\ddots\quad
&\vdots \\
\vdots  \quad &\vdots  \quad &\ddots  \quad &\ddots  \quad &0 \\
\overline{b}_{n}\quad  &\ldots  &\ldots  \quad &\overline{b}_2 \quad
&\overline{b}_1 \end{array}\right),
\end{equation}
and vectors $|p_j\rangle (j=1, \dots, n)$ remain intact.

Summarizing, we conclude that soliton matrices (\ref{newform}) and
(\ref{newform1}) are invariant under the triangular Toeplitz transformations
\begin{equation}\label{transform}
(|p_1\rangle,\ldots,|p_n\rangle)=(|\widetilde{p}_1\rangle,\ldots,|\widetilde{p}_n\rangle)
B, \hspace{0.5cm} \left(\begin{array}{c} \langle \overline{p}_1| \\
\vdots \\ \langle \overline{p}_n| \end{array} \right) =
\overline{B}\left(\begin{array}{c} \langle \widetilde{\overline{p}}_1|
\\ \vdots \\ \langle \widetilde{\overline{p}}_n|
\end{array}\right),
\end{equation}
of the independent vectors $|p_j\rangle$ and $\langle \overline{p}_j|\; (1\le
j\le n)$. Here $B$ and $\overline{B}$ are arbitrary lower and upper triangular
Toeplitz matrices, respectively, in general $(x,t)$-dependent.  Here we point
out that the invariance transformation found in \cite{Nathalie} is given by
$b_j=\overline{b}_j=0 \; (2\le j \le n-1)$, i.e.,  only $b_1, b_n,
\overline{b}_1$ and $\overline{b}_n$ being non-zero. Thus it is just a special
case of the invariance property of the soliton matrices.

The invariance transformations indicate that {\em arbitrary} sets of vectors
$|p_1\rangle,\ldots,|p_n\rangle$ and $ \langle \overline{p}_1|,\ldots,\langle
\overline{p}_n|$ satisfying equations (\ref{S14}) and (\ref{S15}) can be chosen
as the independent vector parameters. This is, in fact, also a {\em necessary
condition} for such vector parameterization of the soliton matrix to be
self-consistent.

Now let us derive the $(x,t)$-dependence of the vector parameters which
enter the soliton matrix. We can start with the fact that the soliton
matrix $\Gamma(k,x,t)$ must satisfy equations (\ref{RH3})-(\ref{RH4})
with some potentials $U(k,x,t)$ and $V(k,x,t)$:
\begin{mathletters}
\label{S27}\begin{eqnarray}
\partial_x \Gamma(k,x,t) &=& \Gamma(k,x,t)\Lambda(k) +
U(k,x,t)\Gamma(k,x,t),
\label{S27a}\\
\partial_t \Gamma(k,x,t) &=& \Gamma(k,x,t)\Omega(k) +
V(k,x,t)\Gamma(k,x,t).
\label{S27b}\end{eqnarray}
\end{mathletters}
The derivation is based on the use of equations (\ref{S14}) and (\ref{S15})
(quite similar to the derivation of equations (\ref{RH8}) in section
\ref{secRH}). First of all we need to find the equations for the triangular
block Toeplitz  matrices ${\bf\Gamma}$ and $\overline{\bf\Gamma}$. To this goal
one needs to differentiate equations (\ref{S27}) with respect to $k$ up to the
$(n-1)$-th order. It is easy to see that, for instance, the equations for the
${\bf\Gamma}$ have the same form as equations (\ref{S27}):
\begin{mathletters}
\label{S28}
\begin{eqnarray}
\partial_x {\bf\Gamma}(k,x,t) &=& {\bf\Gamma}(k,x,t){\bf\Lambda}(k) +
{\bf U}(k,x,t){\bf\Gamma}(k,x,t),
\label{S28a}\\
\partial_t {\bf\Gamma}(k,x,t) &=& {\bf\Gamma}(k,x,t){\bf\Omega}(k) +
{\bf V}(k,x,t){\bf\Gamma}(k,x,t),
\label{S28b}\end{eqnarray}
\end{mathletters}
if we introduce the lower-triangular block Toeplitz  matrices ${\bf\Lambda}$,
${\bf\Omega}$, ${\bf U}$, and ${\bf V}$:
\begin{equation}
{\bf\Lambda} \equiv \left(\begin{array}{cccc}
\Lambda&0 & \ldots& \quad 0\\
\frac{1}{1!}\frac{\text{d}}{\text{d}k}\Lambda
&\ddots&\ddots&\quad\vdots\\
\vdots &\ddots&\Lambda&\quad0
\\
\frac{1}{(n-1)!}\frac{\text{d}^{n-1}}{\text{d}k^{n-1}}
\Lambda&\ldots&\frac{1}{1!}\frac{\text{d}}{\text{d}k}\Lambda&\quad\Lambda
\end{array}\right),\quad
{\bf\Omega} \equiv \left(\begin{array}{cccc}
\Omega&0 & \ldots&\quad 0\\
\frac{1}{1!}\frac{\text{d}}{\text{d}k}\Omega
&\ddots&\ddots&\quad\vdots\\
\vdots &\ddots&\Omega&\quad0
\\
\frac{1}{(n-1)!}\frac{\text{d}^{n-1}}{\text{d}k^{n-1}}
\Omega&\ldots&\frac{1}{1!}\frac{\text{d}}{\text{d}k}\Omega&\quad\Omega
\end{array}\right),
\label{S29}\end{equation}
\begin{equation}
{\bf U} \equiv \left(\begin{array}{cccc}
 U&0 & \ldots& \quad 0\\
\frac{1}{1!}\frac{\text{d}}{\text{d}k} U
&\ddots&\ddots&\quad\vdots\\
\vdots &\ddots& U&\quad 0
\\
\frac{1}{(n-1)!}\frac{\text{d}^{n-1}}{\text{d}k^{n-1}}
 U&\ldots&\frac{1}{1!}\frac{\text{d}}{\text{d}k} U&\quad U
\end{array}\right),\quad
{\bf V} \equiv \left(\begin{array}{cccc}
 V&0 & \ldots&\quad 0\\
\frac{1}{1!}\frac{\text{d}}{\text{d}k} V
&\ddots&\ddots&\quad\vdots\\
\vdots &\ddots& V&\quad0
\\
\frac{1}{(n-1)!}\frac{\text{d}^{n-1}}{\text{d}k^{n-1}}
 V&\ldots&\frac{1}{1!}\frac{\text{d}}{\text{d}k} V&\quad V
\end{array}\right).
\end{equation}
Indeed, this is due to the fact that the matrix multiplication in
(\ref{S28}) exactly reproduces the Leibniz rule for higher-order
derivatives of a product. Similarly, using the equations for
$\Gamma^{-1}$, one finds that
\begin{mathletters}
\label{S30}
\begin{eqnarray}
\partial_x \overline{\bf\Gamma}(k,x,t) &=&
-\overline{\bf\Lambda}(k)\overline{\bf\Gamma}(k,x,t)
 - \overline{\bf\Gamma}(k,x,t)\overline{\bf U}(k,x,t),
\label{S30a}\\
\partial_t \overline{\bf\Gamma}(k,x,t) &=&
-\overline{\bf\Omega}(k)\overline{\bf\Gamma}(k,x,t)
 - \overline{\bf\Gamma}(k,x,t)\overline{\bf V}(k,x,t),
\label{S30b}\end{eqnarray}
\end{mathletters}
for the upper-triangular block Toeplitz matrices $\overline{\bf\Lambda}$,
$\overline{\bf\Omega}$, $\overline{\bf U}$, and $\overline{\bf V}$:
\begin{equation}
\overline{\bf\Lambda} = \left(\begin{array}{cccc}
\Lambda\;&\frac{1}{1!}\frac{\text{d}}{\text{d}k}\Lambda&\ldots&
\frac{1}{(n-1)!}\frac{\text{d}^{n-1}}{\text{d}k^{n-1}}\Lambda\\
0\;
&\Lambda&\ddots& \vdots \\
\vdots\; &\ddots&\ddots&\frac{1}{1!}\frac{\text{d}}{\text{d}k}\Lambda
\\
0\; &\ldots&0 & \Lambda
\end{array}\right),\quad
\overline{\bf\Omega} = \left(\begin{array}{cccc} \Omega\;
&\frac{1}{1!}\frac{\text{d}}{\text{d}k}\Omega&\ldots&
\frac{1}{(n-1)!}\frac{\text{d}^{n-1}}{\text{d}k^{n-1}}\Omega\\
0\;
&\Omega&\ddots& \vdots \\
\vdots\; &\ddots&\ddots&\frac{1}{1!}\frac{\text{d}}{\text{d}k}\Omega \\
0\; &\ldots & 0  &\Omega
\end{array}\right),
\label{S31}\end{equation}
\begin{equation}
\overline{\bf U} = \left(\begin{array}{cccc}
 U\;&\frac{1}{1!}\frac{\text{d}}{\text{d}k} U&\ldots&
\frac{1}{(n-1)!}\frac{\text{d}^{n-1}}{\text{d}k^{n-1}} U\\
0\;
& U&\ddots& \vdots \\
\vdots\; &\ddots&\ddots&\frac{1}{1!}\frac{\text{d}}{\text{d}k} U\\
0\; &\ldots&0 &  U
\end{array}\right),\quad
\overline{\bf V} = \left(\begin{array}{cccc}
 V\; &\frac{1}{1!}\frac{\text{d}}{\text{d}k} V&\ldots&
\frac{1}{(n-1)!}\frac{\text{d}^{n-1}}{\text{d}k^{n-1}} V\\
0\;
& V&\ddots&\vdots \\
\vdots\; &\ddots&\ddots&\frac{1}{1!}\frac{\text{d}}{\text{d}k} V\\
0\; &\ldots &0 &  V
\end{array}\right).
\end{equation}
The $(x,t)$-dependence of the vector parameters
$|p_1\rangle,\ldots,|p_n\rangle$ and $ \langle \overline{p}_1|,\ldots,\langle
\overline{p}_n|$ can be found by differentiation of equations (\ref{S14}) and
(\ref{S15}) with the help of equations (\ref{S28}) and (\ref{S30}). First, we
note that for commuting matrices  the corresponding block Toeplitz matrices as
introduced above also commute with each other. Second, it is shown in the
Appendix that for a diagonal matrix [e.g. $\Lambda(k)x+\Omega(k)t\,$] the
operation of raising to the exponent commutes with the construction of the
block Toeplitz matrix. Therefore, taking into account the invariance property,
we find the $(x,t)$-dependence of the vector parameters as
\begin{mathletters}
\label{S32}
\begin{equation}
\left(\begin{array}{c} |p_1\rangle \\ \vdots \\
|p_n\rangle\end{array}\right) = \exp\left\{{-\bf\Lambda}(k_1)x
-{\bf\Omega}(k_1)t\right\}
\left(\begin{array}{c} |p^{(0)}_1\rangle \\ \vdots \\
|p^{(0)}_n\rangle\end{array}\right),
\label{S32a}\end{equation}
\begin{equation}
(\langle \overline{p}_1|,\ldots,\langle \overline{p}_n|) = (\langle
\overline{p}^{(0)}_1|,\ldots,\langle \overline{p}^{(0)}_n|)
\exp\left\{\overline{\bf\Lambda}(\overline{k}_1)x
+\overline{\bf\Omega}(\overline{k}_1)t\right\}.
\label{S32b}\end{equation}
\end{mathletters}
Here the superscript ``0''  is used to denote constant vectors and the
exponents stand for the triangular block Toeplitz matrices:
\begin{mathletters}
\label{S33}
\begin{equation}
\exp\left\{{-\bf\Lambda}(k_1)x -{\bf\Omega}(k_1)t\right\} =
\left(\begin{array}{cccc}
E(k_1)&0 & \ldots&\; 0\\
\frac{1}{1!}\frac{\text{d}}{\text{d}k}E(k_1)
& \ddots&\ddots&\;\vdots\\
\vdots &\ddots&E(k_1)&\;
0 \\
\frac{1}{(n-1)!}\frac{\text{d}^{n-1}}{\text{d}k^{n-1}}E(k_1)&\ldots&
\frac{1}{1!}\frac{\text{d}}{\text{d}k}E(k_1)&\;
E(k_1)\end{array}\right),
\label{S33a}\end{equation}

\begin{equation}
\exp\left\{\overline{\bf\Lambda}(\overline{k}_1)x
+\overline{\bf\Omega}(\overline{k}_1)t\right\}
 = \left(\begin{array}{cccc}
E^{-1}(\overline{k}_1)&\frac{1}{1!}\frac{\text{d}}{\text{d}k}E^{-1}(\overline{k}_1)&\ldots&
\frac{1}{(n-1)!}\frac{\text{d}^{n-1}}{\text{d}k^{n-1}}E^{-1}(\overline{k}_1)\\
0&E^{-1}(\overline{k}_1)&\ddots&
\vdots \\
\vdots&\ddots&\ddots&\frac{1}{1!}\frac{\text{d}}{\text{d}k}E^{-1}(\overline{k}_1)\\
0&\ldots&0 &E^{-1}(\overline{k}_1)
\end{array}\right),
\label{S33b} \end{equation}
\end{mathletters}
where $E(k) \equiv \exp\left\{-\Lambda(k)x -\Omega(k)t\right\}$. After
the temporal and spatial evolutions for vectors $|p_j\rangle$ and
$\langle \overline{p}_j|$ have been obtained as above, the
corresponding higher-order soliton solution can be obtained from
equations (\ref{RH5}), (\ref{UV3wave}), (\ref{S8a}) and (\ref{S18}).

\section{Application to the three wave interaction  model}
\label{applica}

Here we apply the theory developed in the previous section to the
three-wave interaction model (\ref{3wave}). The three-wave model has
wide applications in nonlinear physics. For instance, under the
additional constraint $u_j =iq_j$ where $q_j$ are real variables, it
describes the ``exact resonance'' in parametric interaction of three
wave packets, while under the reduction of the dispersion laws
(\ref{disper}): $a_3=-a_1$, $a_2=0$, $b_3=-b_1$, $b_2=0$ and the
condition  $u_2=-u_1$, it models the generation of second harmonics.
The usual (fundamental) soliton solutions to the three-wave interaction
model have been well studied (consult Ref.~\cite{NMPZ84}). Such
solitons approach sech profiles as $t \to \pm \infty$ on the
characteristics $x-v_jt=const$.

Let us consider the simplest higher-order solitons in the three-wave
system: solitons which correspond to an elementary higher-order zero of
order 2. Here we should take into account the involution property given
by equation (\ref{invol}). For instance, we have
\[
\overline{k} = k^*,\quad \langle \overline{p}_j| = |p_j\rangle^\dagger
\]
(here and below the overline is associated with the Hermitian or, in
the case of scalar quantities, complex conjugation). Then the soliton
matrix reads
\begin{equation}
\Gamma(k) =  I - (|p_1\rangle,|p_2\rangle)\overline{\cal
K}^{-1}\left(\begin{array}{c} \langle \overline{Z}_2(k)| \\ \langle
\overline{Z}_1(k)|
\end{array}\right),
\label{W1}\end{equation}
where
\begin{equation}
\overline{\cal K} = \left(\begin{array}{cc} \langle
\overline{Z}_2(k_1)|p_1\rangle&\langle \overline{Z}_2(k_1)|p_2\rangle +
\frac{\text{d}}{\text{d}k}
\langle \overline{Z}_2(k_1)|p_1\rangle \\
\langle \overline{Z}_1(k_1)|p_1\rangle&\langle
\overline{Z}_1(k_1)|p_2\rangle + \frac{\text{d}}{\text{d}k} \langle
\overline{Z}_1(k_1)|p_1\rangle\end{array}\right),
\label{W2}\end{equation}
and
\[
\langle \overline{Z}_2(k)| = \frac{\langle
\overline{p}_1|}{k-\overline{k}_1},\quad \langle \overline{Z}_1(k)| =
\frac{\langle \overline{p}_2|}{k-\overline{k}_1} + \frac{\langle
\overline{p}_1|}{(k-\overline{k}_1)^2}.
\]
The $(x,t)$-dependence of the vector parameters
$|p_1\rangle,|p_2\rangle$ has the following form
\begin{equation}
\left(\begin{array}{c} |p_1\rangle \\ |p_2\rangle\end{array}\right) =
\left(\begin{array}{cc} E(k_1) & 0 \\ \frac{\text{d}}{\text{d} k}E(k_1)
& E(k_1)
\end{array}\right)\left(\begin{array}{c} |p^{(0)}_1\rangle \\
|p^{(0)}_2\rangle\end{array}\right), \quad E(k_1) = e^{-ik_1(Ax+Bt)}.
\label{W3}\end{equation}
We denote $k_1 = \xi +i\eta$, where $\xi$ and $\eta$ are real numbers
($\eta>0$ since $k_1$ lies the upper half plane of the complex plane),
and choose the following parameterization of the constant vectors
$|p^{(0)}_1\rangle$ and $|p^{(0)}_2\rangle$:
\begin{equation}
|p^{(0)}_1\rangle = 2i\eta\left(\begin{array}{c}\theta^{(1)}_1
\\\theta^{(1)}_{2} \\\theta^{(1)}_{3}
\end{array}\right) ,\quad
|p^{(0)}_2\rangle = \left(\begin{array}{c}\theta^{(2)}_{1}
\\\theta^{(2)}_{2} \\\theta^{(2)}_{3}
\end{array}\right),
\label{W4}\end{equation}
where $\theta_j^{(i)}$'s are complex constants. It is noted that due to
the invariance property (\ref{transform}), where the matrix $B$
contains two arbitrary complex constants,  we have 2 free components in
each vector in formula (\ref{W4}). Hence there are 10 free real
parameters (including $\xi$ and $\eta$) in the higher-order soliton
solution.

The $(x,t)$-dependence of the components of the vector parameters reads
\begin{equation}
p_{1j} = 2i\eta\theta^{(1)}_je^{f_j/2-i\chi_j},\quad p_{2j} =
\left[\theta^{(2)}_j+f_j\theta^{(1)}_j\right]e^{f_j/2-i\chi_j},
\label{W5}\end{equation}
where
\begin{equation} \label{fchi}
f_j = 2\eta(a_jx+b_jt), \hspace{0.3cm} \chi_j = \xi(a_jx+b_jt), \;\;\;
j=1, 2, 3.
\end{equation}
By simple calculations we obtain the elements of matrix $\overline{\cal
K}$ as
\begin{equation} \label{K12}
\overline{\cal K}_{11} =
-2i\eta\sum_{j=1}^3|\theta^{(1)}_{j}|^2e^{f_j},\quad \overline{\cal
K}_{12} =
-\sum_{j=1}^3\left(\overline{\theta}^{(1)}_{j}\theta^{(2)}_{j}
+(f_j-1)|\theta^{(1)}_{j}|^2\right)e^{f_j},
\end{equation}
\begin{equation} \label{K34}
\overline{\cal K}_{21} =
\sum_{j=1}^3\left(\theta^{(1)}_{j}\overline{\theta}^{(2)}_{j}
+(f_j-1)|\theta^{(1)}_{j}|^2\right)e^{f_j},\quad \overline{\cal K}_{22}
= \frac{1}{2i\eta}\sum_{j=1}^3\left(|\theta^{(2)}_{j}
+(f_j-1)\theta^{(1)}_{j}|^2+|\theta^{(1)}_{j}|^2\right)e^{f_j}.
\end{equation}
It is easy to verify that the determinant of $\overline{\cal K}$ is
\begin{equation}
\det\overline{\cal K} = -
\sum_{i,j=1}^3\left(|\theta^{(1)}_{i}\theta^{(1)}_{j}|^2
+\frac{1}{2}|\theta^{(2)}_{i}\theta^{(1)}_{j}-\theta^{(1)}_{i}\theta^{(2)}_{j}
+(f_i-f_j)\theta^{(1)}_{i}\theta^{(1)}_{j}|^2\right)e^{f_i+f_j},
\label{W6}\end{equation}
which is always non-zero:

For the soliton solution corresponding to the matrix (\ref{W1}) we need
the first-order  term of its asymptotics as $k\to\infty$:
\begin{equation}
\Gamma^{(1)} = -\frac{1}{\det\overline{\cal K}}\left( \overline{\cal
K}_{22} |p_1\rangle \langle \overline{p}_1| +\overline{\cal K}_{11}
|p_2\rangle \langle \overline{p}_2| - \overline{\cal K}_{12}
|p_1\rangle \langle \overline{p}_2| -\overline{\cal K}_{21} |p_2\rangle
\langle \overline{p}_1|\right).
\label{W7}\end{equation}
Using formulae (\ref{W5}) for $|p_1\rangle$ and $|p_2\rangle$ and
(\ref{K12}, \ref{K34}) for the elements of $\overline{\cal K}$ we get
\begin{equation}
\Gamma^{(1)}_{lm} = -\frac{2i\eta}{\det\overline{\cal K}}\,
e^{(f_l+f_m)/2-i(\chi_l-\chi_m)} \sum_{j=1}^3C_{lmj}e^{f_j},
\label{W8}
\end{equation}
where
\[
C_{lmj} = \Bigl[\theta^{(1)}_j\theta^{(1)}_l\left(f_j-f_l-1\right)
+\theta^{(2)}_j\theta^{(1)}_l-\theta^{(2)}_l\theta^{(1)}_j\Bigr]
\Bigl[\overline{\theta}^{(1)}_m\overline{\theta}^{(1)}_j\left(f_m-f_j+1\right)
+\overline{\theta}^{(2)}_m\overline{\theta}^{(1)}_j
-\overline{\theta}^{(2)}_j\overline{\theta}^{(1)}_m\Bigr]
\]

\begin{equation}
- \theta^{(1)}_l\overline{\theta}^{(1)}_m|\theta^{(1)}_j|^2.
\label{W9}
\end{equation}
The three nonlinear waves $u_1,u_2,u_3$ are given by formula
(\ref{u1u2u3}). Thus
\begin{equation}
u_1 = \sqrt{a_1-a_2}\,\Gamma^{(1)}_{12},\quad u_2 =
\sqrt{a_2-a_3}\,\Gamma^{(1)}_{23},\quad u_3 =
\sqrt{a_1-a_3}\,\Gamma^{(1)}_{13},
\label{W12}\end{equation}
where $\Gamma^{(1)}_{ij}$ are given by Eqs. (\ref{W8}) and (\ref{W9}).
To be explicit, our soliton solution corresponding to an elementary
higher-order zero of order 2 in the three-wave interaction system is
\begin{equation} \label{u1}
u_1=-\frac{2i\eta \sqrt{a_1-a_2}}{\det\overline{\cal
K}}e^{(f_1+f_2)/2-i(\chi_1-\chi_2)} \sum_{j=1}^3C_{12j}e^{f_j},
\end{equation}
\begin{equation} \label{u2}
u_2=-\frac{2i\eta \sqrt{a_2-a_3}}{\det\overline{\cal
K}}e^{(f_2+f_3)/2-i(\chi_2-\chi_3)} \sum_{j=1}^3C_{23j}e^{f_j},
\end{equation}
\begin{equation} \label{u3}
u_3=-\frac{2i\eta \sqrt{a_1-a_3}}{\det\overline{\cal
K}}e^{(f_1+f_3)/2-i(\chi_1-\chi_3)} \sum_{j=1}^3C_{13j}e^{f_j},
\end{equation}
where $\det\overline{\cal K}, C_{ijk}, f_k$ and $\chi_k$ are given by
Eqs. (\ref{fchi}), (\ref{W6}) and (\ref{W9}).

The above solutions are fairly complicated. But some information about
them can be gained from considering the asymptotics as $t\to\pm\infty$.
Evidently, the $t$-asymptotics is nonzero only on the characteristics:
\[
z_1 \equiv (f_1 - f_2)/2 = \eta(a_1-a_2)(x-v_1t), \]
\[
z_2 \equiv (f_2 - f_3)/2 = \eta(a_2-a_3)(x-v_2t),
\]
\[
z_3 \equiv (f_1 - f_3)/2 = \eta(a_1-a_3)(x-v_3t). \] The asymptotic formulae
depend on the relation between the velocities of the waves. For definiteness,
let us choose $v_2<v_1$.  This is equivalent to the condition $\varepsilon>0$
in view that
\[ \varepsilon=\left(\frac{(a_1-a_2)(a_2-a_3)}{a_1-a_3}\right)^{1/2}(v_1-v_2),
\]
as follows from formulae (\ref{add1})-(\ref{add2}).  Then the condition
(\ref{ineqv}) requires that $v_3$ lies between $v_2$ and $v_1$:
\begin{equation}
v_2<v_3<v_1.
\label{W14}\end{equation}
Further, we notice that any solution
$\tilde{u}_1,\tilde{u}_2,\tilde{u}_3$ of the three-wave interaction
model (\ref{3wave}) in the case of the opposite inequality, i.e.
$v_1<v_3<v_2$, is mapped onto the solution satisfying the inequality
(\ref{W14}) by the following transformation:
$\tilde{u}_j(x,t;v_1,v_2,v_3)=-u_j(x,-t;-v_1,-v_2,-v_3)$. Thus, the
case of $v_1<v_2$ is easy to recover (it describes the reverse process
to that of $v_2<v_1$).

The asymptotic formulae also depend on whether some of the components
in vectors $\theta^{(1)}$ and $\theta^{(2)}$ are zero or not. We first
consider the generic case when none of the parameters $\theta^{(1)}_j$
for $j=1,2,3$ is zero. Define the following real quantities:
\begin{eqnarray}
\alpha_{lm} =
\ln\left(\frac{|\theta^{(1)}_m|}{|\theta^{(1)}_l|}\right),\quad
\varrho_{lm}+i\sigma_{lm} = \frac{1}{2}\left(
\frac{\theta^{(2)}_l}{\theta^{(1)}_l} -
\frac{\theta^{(2)}_m}{\theta^{(1)}_m}\right)
+\alpha_{lm},  \nonumber \\
\varphi^{(s)}_j = \text{arg}(\theta^{(s)}_j), \quad \varphi_{lm} =
\varphi^{(1)}_l - \varphi^{(1)}_m -\frac{\xi\alpha_{lm}}{\eta},  \quad
\quad \quad
\label{W19}
\end{eqnarray}
and denote
\begin{equation}
z_{12} = z_1 - \alpha_{12},\quad z_{23} = z_2 - \alpha_{23},\quad
z_{13} = z_3 - \alpha_{13}.
\label{W21}\end{equation}
Then, simple calculations show that the asymptotics of the waves
(\ref{u1}), (\ref{u2}) and (\ref{u3}) are as follows:
\begin{equation}
\label{u123asym1}
u_1 \to 0, \quad u_2 \to 0, \quad  t\to -\infty; \quad \quad u_3 \to 0,
\quad t \to \infty,
\end{equation}
\begin{equation}
\label{u1asym1}
u_1 \rightarrow 2i\eta\sqrt{a_1-a_2}
\frac{(z_{12}+\varrho_{12})\sinh{z_{12}}
-(1+i\sigma_{12})\cosh{z_{12}}}
{\cosh^2{z_{12}}+(z_{12}+\varrho_{12})^2+\sigma_{12}^2}
e^{i(\varphi_{12}-\xi z_{12}/\eta )},  \quad \quad t\to \infty,
\end{equation}
\begin{equation}
\label{u2asym1}
u_2 \rightarrow 2i\eta\sqrt{a_2-a_3}
\frac{(z_{23}+\varrho_{23})\sinh{z_{23}}
-(1+i\sigma_{23})\cosh{z_{23}}}
{\cosh^2{z_{23}}+(z_{23}+\varrho_{23})^2+\sigma_{23}^2}
e^{i(\varphi_{23}-\xi z_{23}/\eta )},  \quad \quad t\to \infty,
\end{equation}
\begin{equation}
\label{u3asym1}
u_3\rightarrow 2i\eta\sqrt{a_1-a_3}
\frac{(z_{13}+\varrho_{13})\sinh{z_{13}}
-(1+i\sigma_{13})\cosh{z_{13}}}
{\cosh^2{z_{13}}+(z_{13}+\varrho_{13})^2+\sigma_{13}^2}
e^{i(\varphi_{13}-\xi z_{13}/\eta )}, \quad \quad t\to -\infty.
\end{equation}
We see that as $t\to-\infty$, only the pumping wave $u_3$ is non-zero,
while as $t\to\infty$ only the elementary waves $u_1$ and $u_2$ are
nonzero. Thus in the generic case of higher-order solitons under
condition (\ref{W14}), the solution describes the breakdown of the
pumping higher-order soliton $u_3$ into the higher-order solitons of
the elementary waves $u_1$ and $u_2$. [For the opposite inequalities in
formula (\ref{W14}), the solution describes the reverse process: merger
of the two elementary waves $u_1$ and $u_2$ into the pumping wave
$u_3$.] These properties are identical to fundamental solitons (see,
for instance, pp. 174-184 in Ref.~\cite{NMPZ84}). However, differences
between higher-order solitons and fundamental solitons are also
obvious: none of the asymptotics (\ref{u1asym1}) to (\ref{u3asym1}) of
the higher-order solitons is sech-shaped, while the asymptotics of
fundamental solitons are all sech-shaped.

Asymptotics (\ref{u123asym1}) to (\ref{u3asym1}) are invalid in the
non-generic cases when at least one of the parameters $\theta^{(1)}_j$,
$j=1,2,3$, is zero. Consider first the case of $\theta^{(1)}_1=0$ and
$\theta^{(1)}_2$ and $\theta^{(1)}_3$ being non-zero. There are two
possibilities depending on whether $\theta^{(2)}_1$ is zero or not.

(a) If $\theta^{(2)}_1 \ne 0$, then the asymptotics of the waves
(\ref{u1}), (\ref{u2}) and (\ref{u3}) become
\begin{equation}
u_1 \to 0, \quad t\to -\infty; \quad \quad u_3 \to 0, \quad t\to
\infty;
\end{equation}
\begin{equation}
u_1 \to -i\eta\sqrt{a_1-a_2}e^{i(\varphi^{(2)}_1-\varphi^{(1)}_2-\xi
z_1/\eta)} \text{sech}(z_1-\beta_1), \quad t\to\infty,
\end{equation}
\begin{equation}
u_2 \to -i\eta\sqrt{a_2-a_3}e^{i(\varphi^{(1)}_2-\varphi^{(1)}_3-\xi
z_2/\eta)} \text{sech}(z_2-\beta_2), \quad t\to -\infty,
\end{equation}
\begin{equation}
u_2 \rightarrow 2i\eta\sqrt{a_2-a_3}
\frac{(z_{23}+\varrho_{23})\sinh{z_{23}}
-(1+i\sigma_{23})\cosh{z_{23}}}
{\cosh^2{z_{23}}+(z_{23}+\varrho_{23})^2+\sigma_{23}^2}
e^{i(\varphi_{23}-\xi z_{23}/\eta )},  \quad \quad t\to \infty,
\end{equation}
\begin{equation}
u_3 \rightarrow
-i\eta\sqrt{a_1-a_3}e^{i(\varphi^{(2)}_1-\varphi^{(1)}_3-\xi z_3/\eta)}
\text{sech}(z_3-\beta_3), \quad t\to-\infty,
\label{W24}
\end{equation}
where parameters $\beta_{j} \; (j=1, 2, 3)$ are defined as
\begin{equation} \label{beta}
\beta_1
=\ln\left(\frac{|\theta^{(1)}_2|}{|\theta^{(2)}_1|}\right),\quad
\beta_2
=\ln\left(\frac{|\theta^{(1)}_3|}{|\theta^{(1)}_2|}\right),\quad
\beta_3 =\ln\left(\frac{|\theta^{(1)}_3|}{|\theta^{(2)}_1|}\right).
\label{betas123}\end{equation} The above asymptotics have two important
features. One is that as $t\to -\infty$, both $u_3$ and $u_2$ waves are
non-zero. Thus, the higher-order soliton solution (\ref{u1}) to (\ref{u3}) in
this non-generic case does not describe the breakdown of the pumping wave
$u_3$. Instead, it describes a {\em new process}:
\begin{equation}
u_2+u_3 \to u_1+u_2.
\label{newpr1}\end{equation}
This is very different from fundamental solitons. The other feature is
that the waves $u_2$ and $u_3$ as $t\to -\infty$ and the wave  $u_1$ as
$t\to \infty$ all have sech profiles, but the wave $u_2$  as $t\to
\infty$ is the higher-order soliton.

(b) If $\theta^{(2)}_1 =0$, then the higher-order solitons (\ref{u1})
to (\ref{u3}) are degenerate:
\begin{equation}
\label{trivial}
u_1(x, t)=u_3(x, t)=0 \quad u_2(x, t)=u_2^{(0)}(x-v_2 t),
\end{equation}
where $u_2^{(0)}(x)$ is the initial solution of $u_2$. This is a
trivial solution.

If two components of the vector $\theta^{(1)}$ are zero, then
higher-order soliton solutions (\ref{u1}) to (\ref{u3}) reduce to
fundamental-soliton solutions or trivial solutions. For instance, if
$\theta^{(1)}_1=\theta^{(1)}_2=0$, $\theta^{(1)}_3\ne 0$,
$\theta^{(2)}_1\ne 0$, and $\theta^{(2)}_2\ne 0$, then the asymptotics
of the waves become
\begin{equation}
u_1 \to 0, \quad u_2 \to 0, \quad t \to -\infty; \quad \quad u_3 \to 0,
\quad t \to \infty;
\end{equation}
\begin{equation}
u_1 \to -i\eta\sqrt{a_1-a_2}e^{i(\varphi^{(2)}_1-\varphi^{(2)}_2-\xi
z_1/\eta)} \text{sech}(z_1-\tilde{\beta}_1), \quad t\to\infty,
\end{equation}
\begin{equation}
u_2 \to -i\eta\sqrt{a_2-a_3}e^{i(\varphi^{(2)}_2-\varphi^{(1)}_3-\xi
z_2/\eta)} \text{sech}(z_2-\tilde{\beta}_2), \quad t\to\infty,
\end{equation}
\begin{equation}
u_3 \rightarrow
-i\eta\sqrt{a_1-a_3}e^{i(\varphi^{(2)}_1-\varphi^{(1)}_3-\xi z_3/\eta)}
\text{sech}(z_3-\beta_3), \quad t\to-\infty,
\end{equation}
where $\beta_3$ is defined in equation (\ref{betas123}) and
\begin{equation}
\tilde{\beta}_1=\ln\left(\frac{|\theta^{(2)}_2|}{|\theta^{(2)}_1|}\right),\quad
\tilde{\beta}_2=\ln\left(\frac{|\theta^{(1)}_3|}{|\theta^{(2)}_2|}\right).
\label{tilded}
\end{equation}
This is the fundamental soliton solution. If one or both of
$\theta^{(2)}_1$ and $\theta^{(2)}_2$ is zero, the solution is trivial
(similar to (\ref{trivial})). We note that when
$\theta^{(1)}_1=\theta^{(1)}_2=0$, then $\theta^{(1)}_3$ can not be
zero, because otherwise, the denominator $\det\overline{\cal K}$ in the
solution is zero.

It turns out that consideration of the case when $\theta^{(1)}_3=0$ is
similar to the above case of $\theta^{(1)}_1=0$ with the only
difference that now the elementary waves $u_1$ and $u_2$ are
interchanged. For instance, when $\theta^{(1)}_3=0$,
$\theta^{(1)}_1\ne0$, $\theta^{(1)}_2\ne0$, and $\theta^{(2)}_3\ne0$ we
have the following (also new) process:
\begin{equation}
u_1+u_3 \to u_1 +u_2,
\label{newpr2}\end{equation}
where the waves $u_1$ and $u_3$ as $t\to -\infty$ and the wave $u_2$ as
$t\to \infty$ all have sech profiles, while  the wave $u_1$ as $t\to
\infty$ is the higher-order soliton [see equation (\ref{u2asym1})].

The only (different) case which is left to consider is the case of
$\theta^{(1)}_2=0$ with $\theta^{(1)}_1\ne0$ and $\theta^{(1)}_3\ne0$.
The asymptotics depends on whether $\theta^{(2)}_2$ is zero or not. In
the former case, i.e. $\theta^{(2)}_2=0$, we have a degenerate
solution,
\[
u_1(x,t) = u_2(x,t) = 0, \quad u_3(x,t)=u_3^{(0)}(x-v_3t),
\]
which is similar to solution (\ref{trivial}). If however
$\theta^{(2)}_2\ne0$, then the asymptotics of the waves
(\ref{u1})-(\ref{u3}) are as follows:
\begin{equation}
u_1\to0,\quad u_2\to0,\quad t\to-\infty,
\end{equation}
\begin{equation}
u_3 \to 2i\eta\sqrt{a_1-a_3} \frac{(z_{13}+\varrho_{13})\sinh{z_{13}}
-(1+i\sigma_{13})\cosh{z_{13}}}
{\cosh^2{z_{13}}+(z_{13}+\varrho_{13})^2+\sigma_{13}^2}
e^{i(\varphi_{13}-\xi z_{13}/\eta )}, \quad t\to -\infty;
\end{equation}
\begin{equation}
u_1 \to -i\eta\sqrt{a_1-a_2}e^{i(\varphi^{(1)}_1-\varphi^{(2)}_2-\xi
z_1/\eta)} \text{sech}(z_1-\hat{\beta}_1), \quad t\to\infty,
\end{equation}
\begin{equation}
u_2 \to -i\eta\sqrt{a_2-a_3}e^{i(\varphi^{(2)}_2-\varphi^{(1)}_3-\xi
z_2/\eta)} \text{sech}(z_2-\tilde{\beta}_2), \quad t\to\infty,
\end{equation}
\begin{equation}
u_3 \to -i\eta\sqrt{a_1-a_3}e^{i(\varphi^{(1)}_1-\varphi^{(1)}_3-\xi
z_3/\eta)} \text{sech}(z_3-\hat{\beta}_3), \quad t\to\infty,
\end{equation}
where $\tilde\beta_2$ is defined in equation (\ref{tilded}) and
\begin{equation}
\hat\beta_1 =
\ln\left(\frac{|\theta^{(2)}_2|}{|\theta^{(1)}_1|}\right),\quad
\hat\beta_3 =
\ln\left(\frac{|\theta^{(1)}_3|}{|\theta^{(1)}_1|}\right).
\end{equation}
These asymptotic formulae describe yet another new process:
\begin{equation}
u_3\to u_1 + u_2 +u_3,
\label{newpr3}\end{equation}
where waves $u_1$, $u_2$, and $u_3$ as $t\to \infty$ all have sech
profiles, while  the pumping wave $u_3$ as $t\to -\infty$ is more
complicated. Thus, this process describes a  breakup of the
higher-order pumping wave into three sech waves, the two elementary
waves and the pumping wave.


Lastly, we present the graphical pictures of the above higher-order
solitons for both the generic and non-generic cases. In all figures,
the common solution parameters are $(a_1, a_2, a_3)=(2, 1, -1)$, $(b_1,
b_2, b_3)=(-0.5, 2, 1),$ $\xi=1, \eta=1$, and $\theta^{(2)}=(-1, 1+i,
2)$. Only the vector $\theta^{(1)}$ is different. It is easy to check
that for these parameters, the inequality (\ref{W14}) holds, thus the
asymptotics of these higher-order solitons have been described in the
previous text. In all figures, the solid lines are $|u_1|$, the dashed
lines are $|u_2|$, and the dashed-dotted lines are $|u_3|$.

First, we illustrate the generic solution with $\theta^{(1)}=(1, i, -1)$ in
Fig. 1. As we can see from this figure as well as the asymptotics
(\ref{u123asym1}) to (\ref{u3asym1}), as $t \to -\infty$, only the pumping
$u_3$ solution is non-zero. As $t\to \infty$, this $u_3$ wave breaks into
elementary $u_1$ and $u_2$ waves. This process is similar to fundamental
solitons. But there is a difference: the asymptotics of each wave in Fig. 1 has
a complex structure which signals that it is a higher-order soliton instead of
a fundamental soliton. Next, we let $\theta^{(1)}_1$ approach zero.
Specifically, we let $\theta^{(1)}_1=10^{-4}$, while the $\theta^{(1)}_2$ and
$\theta^{(1)}_3$ values remain the same. The corresponding soliton solution is
illustrated in Fig. 2. We see that in this case, the pumping $u_3$ wave at
$t\to -\infty$ splits into two sech pulses. As time moves on, the front $u_3$
sech pulse breaks into $u_1$ and $u_2$ sech pulses. Then this $u_2$ sech pulse
and the back $u_3$ sech pulse interact. The final outcome is two $u_1$ sech
pulses moving in the positive $x$ direction, and a higher-order $u_2$ wave
moving in the negative $x$ direction. Thirdly, we consider the non-generic case
where $\theta^{(1)}_1=0$, while $\theta^{(1)}_2$ and $\theta^{(1)}_3$ still do
not change. This soliton solution is illustrated in Fig. 3. We see that as $t
\to -\infty$, both the $u_3$ and $u_2$ waves are non-zero and sech-shaped.
After their interaction, the pumping $u_3$ wave is depleted, and a new $u_1$
sech wave and a higher-order $u_2$ wave are created. We note that this $u_2+u_3
\to u_1+u_2$ process is novel, and it has not been carefully investigated
before. Fourthly, we consider the non-generic case where $\theta^{(1)}_1=
\theta^{(1)}_2=0$, and $\theta^{(1)}_3$ is still $-1$. This solution is
illustrated in Fig. 4. We see that it is the same as a fundamental soliton
solution, and it describes the process of a pumping $u_3$ sech wave breaking
into two elementary $u_1$ and $u_2$ sech waves. Thus, our higher-order soliton
solution reduces to a fundamental soliton solution as a special case. Lastly,
we consider the non-generic case where $\theta^{(1)}_2=0$ while
$\theta^{(1)}_1=1$ and $\theta^{(1)}_3=-1$ as in Fig. 1. This solution is shown
in Fig. 5. As we can see, as $t\to -\infty$, the only non-zero wave is the
pumping wave $u_3$, which is a higher-order soliton. As $t\to \infty$, this
pumping wave breaks up into a sech waves in each component. Thus, this is the
new $u_3 \to u_1+u_2+u_3$ process which we have presented in the text above.

We conclude this section with some comments on the soliton solutions to
the three-wave model corresponding to the higher-order zeros of order
$n\ge 2$.  If the higher-order zero is elementary, i.e. when the sequence
of ranks in formula (\ref{S7}) is rank$P_j=1$, $j=1,...,n$, the
corresponding soliton solutions can be derived using the soliton matrix
(\ref{newform}). However, there are two other possible sequences of
ranks in formula (\ref{S7}), namely
\[(a)\quad  \mbox{rank}P_j=2, \; j=1,...,r, \quad n=2r;
\hspace{6.8cm}\]

\vspace{-0.6cm}
\[(b) \quad \mbox{rank}P_j=2, \; j=1,...,r; \quad \mbox{rank}P_j=1, \;
j=r+1,...,r+s, \quad n=2r+s.\]

We note that the soliton matrix for the sequence of ranks (a), which
corresponds to the higher-order zero of order $2r$, has an equivalent
soliton matrix  corresponding to the elementary higher-order zero of order
$r$. Indeed, let us consider the soliton matrix in the representation
(\ref{S6}),  where each $\chi_j(k)$ is defined similar to formula
(\ref{S4}) [$P_j$ substituted for $P_1$] with rank$P_j=2$. Consider the
following procedure. First,  define new projectors $Q_j=I-P_j$.
Evidently rank$Q_j=1$. Second, multiply the soliton matrix $\Gamma(k)$
(\ref{S6}) by a scalar quotient,
\begin{equation}
\tilde{\Gamma}(k) =
\left(\frac{k-\overline{k}_1}{k-k_1}\right)^r\Gamma(k),
\label{trick}\end{equation}
such that each $\chi_j(k)$, $j=1,...,r$, gets a multiplier
$(k-\overline{k}_1)/(k-k_1)$. We have
\[
\tilde{\chi}_j(k) \equiv \frac{k-\overline{k}_1}{k-k_1}\chi_j(k)=
\frac{k-\overline{k}_1}{k-k_1}\left(\frac{k-k_1}{k-\overline{k}_1}I
+\frac{k_1-\overline{k}_1}{k-\overline{k}_1}Q_j\right) =
I+\frac{k_1-\overline{k}_1}{k-k_1}Q_j,
\]
and
\begin{equation} \label{gammatilde}
\tilde{\Gamma}(k)=\tilde{\chi}_r(k) \dots \tilde{\chi}_2(k)
\tilde{\chi}_1(k).
\end{equation}
Evidently, the new matrix $\tilde{\Gamma}(k)$ in (\ref{gammatilde})
satisfies the linear system of equations (\ref{S27a})-(\ref{S27b}) for
the original matrix $\Gamma(k)$. Furthermore, it corresponds to an
elementary higher-order zero of order $r$, though now in the complementary
half plane: $k=\overline{k}_1$. It is noted that in this section we
considered a soliton matrix $\Gamma(k)$ corresponding to a zero
$k_1=\xi+i\eta$ lying in the upper half plane, i.e. with $\eta>0$.
However, the case of $\eta<0$ is admissible as well. The only
significant change would be in the asymptotic formulae, and the effect
of this change is similar to reversing the time variable: $t\to-t$.
Thus, the sequence of ranks in case (a) brings no new higher-order
soliton solutions as compared to the simple sequence of ranks, but the
solution process is reversed. For the fundamental soliton solutions, a
similar fact has been noted in Ref.~\cite{NMPZ84}, where it is
mentioned that the fundamental soliton corresponding to the projector
of rank 2 describes the three-wave interaction process which is reverse
to that of the soliton solution corresponding to the projector of rank
1.

There is no transformation similar to (\ref{trick}) for case (b) (similar
multiplication will produce a rational matrix function having poles in both
half planes, thus such a  matrix does  not  belong to the class of soliton
matrices). Higher-order soliton solutions in this case require construction of
the soliton matrices for non-elementary higher-order zeros and will be
addressed in a forthcoming  paper.

\section{Conclusion}
We have proposed a unified and systematic approach to study the higher-order
soliton solutions of nonlinear PDEs integrable by the Riemann-Hilbert problem
of arbitrary matrix dimension. We have derived the soliton dressing matrix for
the  elementary higher-order zeros in the $N\times N$-dimensional spectral
problem, i.e., zeros having the geometric multiplicity 1.  The associated
higher-order solitons in the $N$-wave system have also been obtained.  We have
also clarified that the soliton dressing ansatz proposed in \cite{Nathalie} is
the general soliton matrix for the nonlinear Schr\"odinger equation (where
$N=2$), thus the soliton solutions obtained in \cite{Nathalie} are the most
general higher-order solitons in the the nonlinear Schr\"odinger equation. For
$N\times N$-dimensional spectral problems the soliton dressing ansatz of
\cite{Nathalie} corresponds to the elementary higher-order solitons.

We have applied our theory to the three-wave interaction model, and the
simplest higher-order soliton solution has been obtained. The generic case of
this solution describes the process $u_3 \leftrightarrow  u_1+u_2$, similar to
fundamental solitons. But each wave involved here is higher-order. The
non-generic case of this solution could describe three new processes. The first
two are similar to each other:
\begin{eqnarray*}
& & u_1+u_3\leftrightarrow  u_1+u_2, \\
& & u_2+u_3 \leftrightarrow u_1+u_2.
\end{eqnarray*}
Here the waves on the left are all sech waves; the waves on the right
are a sech wave and a higher-order wave. The third process reads
\[
u_3 \leftrightarrow u_1 +u_2 +u_3,
\]
where the pumping wave on the left is a higher-order wave, and the
waves on the right all have sech shape. The non-generic solutions could
also reduce to fundamental solitons or trivial solutions as special
cases.

We anticipate that the higher-order soliton solutions will have wide
applications. First of all, the new processes they describe may find
physical applications where three-wave interaction takes place. Second,
as it has been mentioned in Ref.~\cite{OptLett}, the higher-order
soliton solution describes a weak bound state of solitons, thus it may
appear in the study of the train propagation of solitons with nearly
equal amplitudes and velocities in nonlinear integrable PDEs. The usual
approach in the analytical study of the soliton trains is reduction of
the governing equations for the soliton parameters to the complex Toda
chain (consult, for instance,
Refs.~\cite{TodaNLS1,TodaNLS2,TodaMNLS,Manakovchain}). The higher-order
soliton approach may provide an alternative to this study. Thirdly,
multi-hump solitary waves in the non-integrable nonlinear PDEs can be
another field of application of the higher-order solitons. For
instance, the so-called multisoliton complexes, or more precisely,
oscillatory and stationary solitons observed in an oscillating water
trough \cite{Wu,Wang1,Wang2a,Wang2b,Wang2c}  and subsequently
reproduced in numerical simulations \cite{Wang2a,Wang2b,Wang2c} of the
governing parametrically driven, damped NLS equation may have the same
relation to the higher-order solitons as the usual solitary-wave
solutions of the non-integrable PDEs to the fundamental solitons.
Analytical study of the soliton complexes needs the perturbation theory
for the higher-order solitons, just as the study of usual solitary-wave
solutions needs the perturbation theory for the fundamental solitons.
The perturbation theory for the higher-order solitons can be developed
in a similar way as it is  done for the fundamental solitons (see for
instance Ref.~\cite{Shch1,YangKaup,Yang00}). Such a theory is left for
future studies.

Lastly, we point out that the soliton matrices for the elementary zeros serve
as the building blocks for the general case of zeros with arbitrary geometric
multiplicity. This work is in progress and will be reported in a forthcoming
paper. There the most general higher-order soliton solution for the $N$-wave
system will be given.

\acknowledgements

Helpful discussions with Professor Mark Ablowitz are gratefully acknowledged.
The work by V.S. was sponsored by the University of Vermont. The warm
hospitality of the Department of Mathematics and Statistics is also
acknowledged with gratitude. The work by J.Y. was supported in part by AFOSR
under contract USAF F49620-99-1-0174 and by NSF under grant DMS-9971712.

\appendix
\section{Exponent of the Toeplitz matrices}

Here we show that for a diagonal matrix $M(k)$ the exponent of the block
Toeplitz matrix, defined as in formula  (\ref{S29}), and the block Toeplitz
matrix of the exponent of $M(k)$  coincide. As the derivatives
$\frac{\text{d}^m}{\text{d}k^m}M(k)$ commute with each other, it is enough to
prove this statement for a scalar function. Consider, for example, the
lower-triangular Toeplitz matrix of a scalar function $f(k)$:

\begin{equation}
{\bf F} = \left( \begin{array}{cccc}
f&0&\ldots&\quad 0\\
\frac{\text{d}}{\text{d}k}f&f&\ldots&\quad 0\\
\vdots&\vdots&\ddots&\quad\vdots\\
\frac{\text{d}^{n-1}}{\text{d}k^{n-1}}f&\frac{\text{d}^{n-2}}{\text{d}k^{n-2}}f&\ldots&\quad
f
\end{array} \right).
\label{A1}\end{equation}
It  can be rewritten in the following form
\begin{equation}
{\bf F} = H_0f + H_{1}\frac{\text{d}f}{\text{d}k} + \ldots +
H_{n-1}\frac{\text{d}^{n-1}f}{\text{d}k^{n-1}},\quad
\left(H_j\right)_{l,m} \equiv \delta_{l+j,m}.
\label{A2}\end{equation}
Note the product rule for the ``diagonals'': $H_jH_i = H_{j+i}$ and
that for $j+i>n-1$ the product  is zero. Therefore, the exponent of
${\bf F}$ is a finite sum of the diagonals~$H_j$:
\begin{equation}
\exp({\bf F}) = c_0 H_0 + c_1 H_1 + \ldots +c_{n-1} H_{n-1},
\label{A3}\end{equation}
where $c_0,\ldots,c_{n-1}$ are constants. Due to the formula $H_j =
H_1^j$, $j=0,\ldots,n-1$, computing the coefficients $c_j$ is
equivalent to taking the finite sum of the first $n$ terms of the
Taylor expansion of an equivalent scalar function:
\begin{equation}
\exp\left\{\sum_{j=0}^{n-1}\frac{\text{d}^j f(k)}{\text{d}k^j}
\epsilon^j\right\} =c_0 + c_1 \epsilon + \ldots +c_{n-1} \epsilon^{n-1}
+ {\cal O}\{\epsilon^{n}\},
\label{A4}\end{equation}
where $\epsilon$ is the parameter of the Taylor expansion which
represents $H_1$. On the other hand, computing the Taylor expansion
reduces to taking derivatives with respect to $k$ of
$\exp\{f(k+\epsilon)\}$ at $\epsilon = 0$:
\begin{eqnarray}
\exp\left\{\sum_{j=0}^{n-1}\frac{\text{d}^j f(k)}{\text{d}k^j}
\epsilon^j\right\} = \exp\left\{f(k+\epsilon)\right\}+{\cal
O}\{\epsilon^{n}\}
\hspace{5.5cm} \nonumber \\
= \exp\{f(k)\} + \frac{1}{1!}\frac{\text{d}}{\text{d}k}\exp\{f(k)\}
\epsilon + \ldots +
\frac{1}{(n-1)!}\frac{\text{d}^{n-1}}{\text{d}k^{n-1}}\exp\{f(k)\}
\epsilon^{n-1} + {\cal O}\{\epsilon^{n}\}.
\label{A5}\end{eqnarray}
Therefore
\begin{equation}
c_j=\frac{1}{j!}\frac{\text{d}^{j}}{\text{d}k^{j}}\exp\{f(k)\}, \quad
j=1, \dots, n-1,
\end{equation}
thus,
\begin{equation}
\exp\{{\bf F}\} = H_0\exp\{f(k)\} + H_1
\frac{1}{1!}\frac{\text{d}}{\text{d}k}\exp\{f(k)\} + \ldots
+H_{n-1}\frac{1}{(n-1)!}\frac{\text{d}^{n-1}}{\text{d}k^{n-1}}\exp\{f(k)\}.
\label{A6}\end{equation}
Q.E.D.

\begin{figure}[p]
\begin{center}
\parbox{5cm}{\postscript{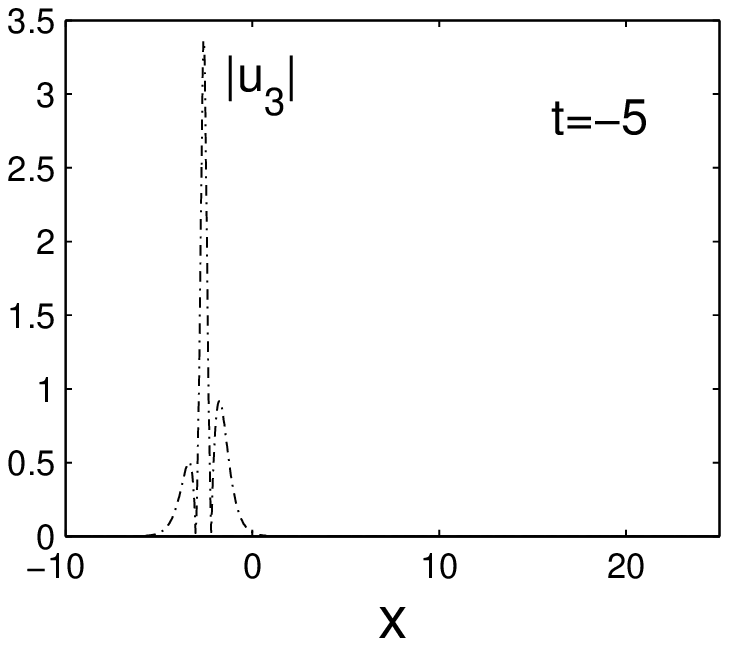}{1.0}} \hspace{0.4cm}
\parbox{5cm}{\postscript{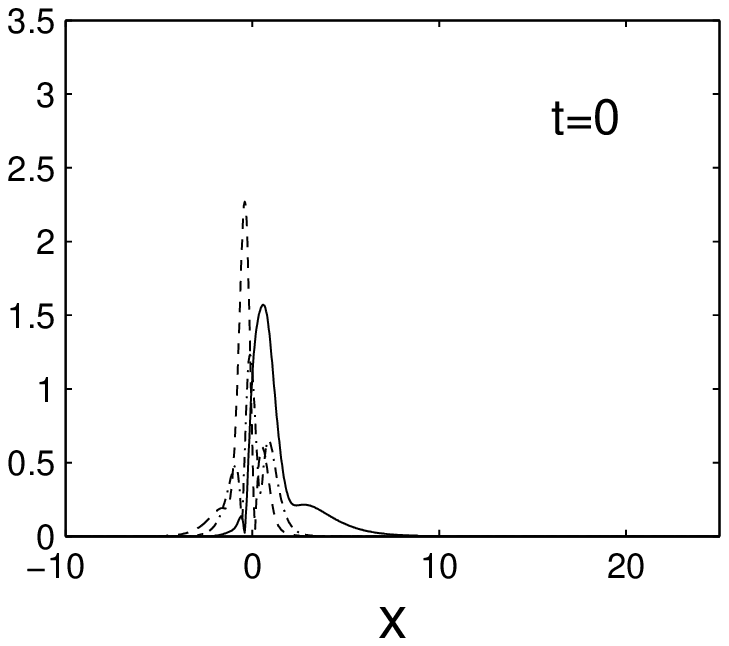}{1.0}} \hspace{0.4cm}
\parbox{5cm}{\postscript{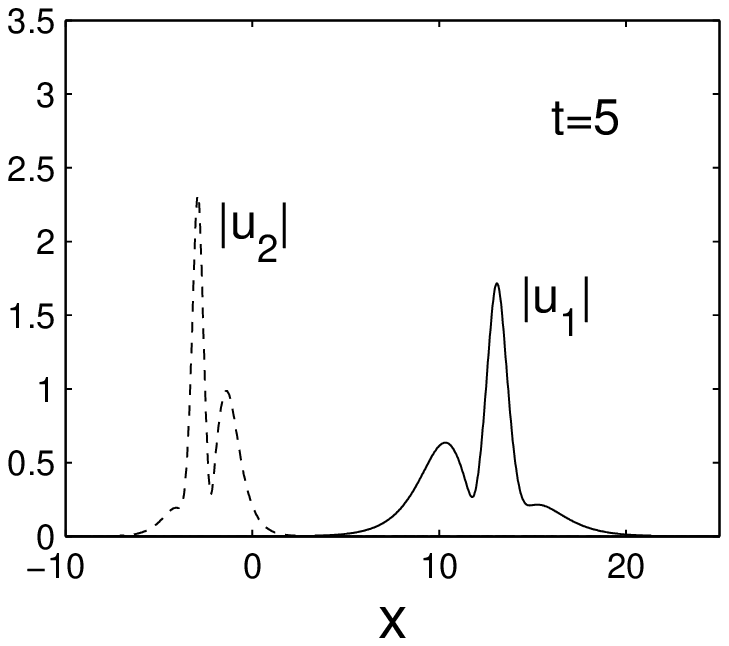}{1.0}}

\vspace{0.5cm} \caption{A generic higher-order soliton solution which
describes the breaking of the higher-order pumping $u_3$ wave into
higher-order elementary $u_1$ and $u_2$ waves, i.e., the $u_3 \to
u_1+u_2$ process. Here, the solution parameters are $(a_1, a_2,
a_3)=(2, 1, -1)$, $(b_1, b_2, b_3)=(-0.5, 2, 1),$ $\xi=1, \eta=1,$
$\theta^{(1)}=(1, i, -1)$ and $\theta^{(2)}=(-1, 1+i, 2)$. In all
figures here and below, solid lines are $|u_1|$, dashed lines are
$|u_2|$, and dash-dotted lines are $|u_3|$.
  }
\end{center}
\end{figure}

\begin{figure}[p]
\begin{center}
\parbox{17cm}{\postscript{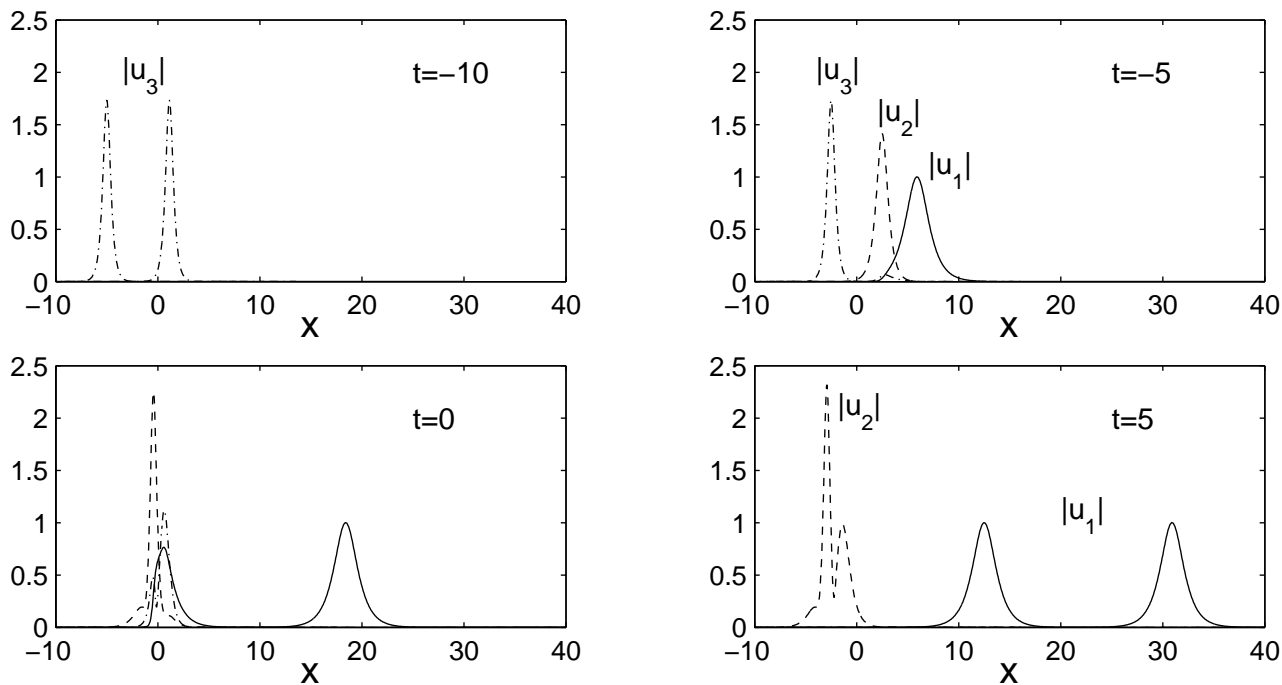}{1.0}}

\vspace{0.5cm} \caption{Another generic higher-order soliton solution
with a very small $\theta^{(1)}_1$ value. Here
$\theta^{(1)}_1=10^{-4}$, while the other solution parameters are the
same as in Fig. 1. }
\end{center}
\end{figure}

\begin{figure}[p]
\begin{center}
\parbox{5cm}{\postscript{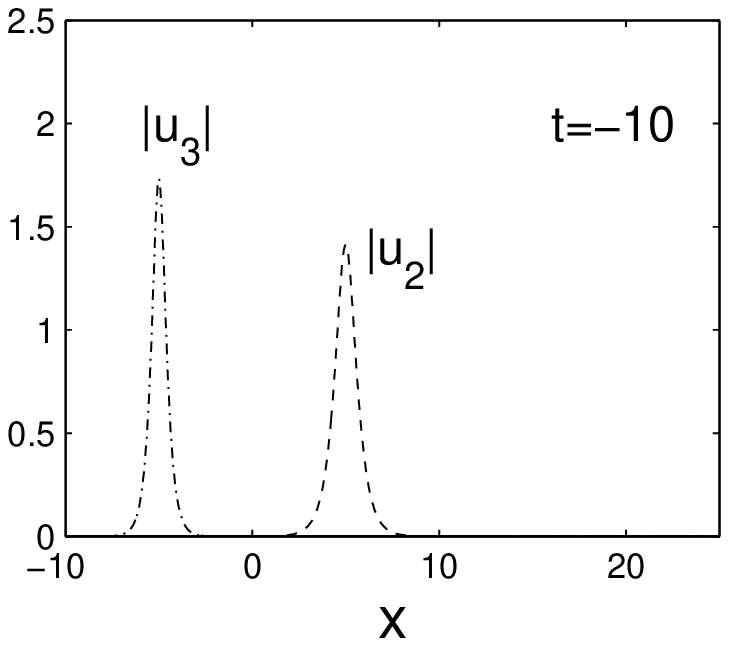}{1.0}} \hspace{0.4cm}
\parbox{5cm}{\postscript{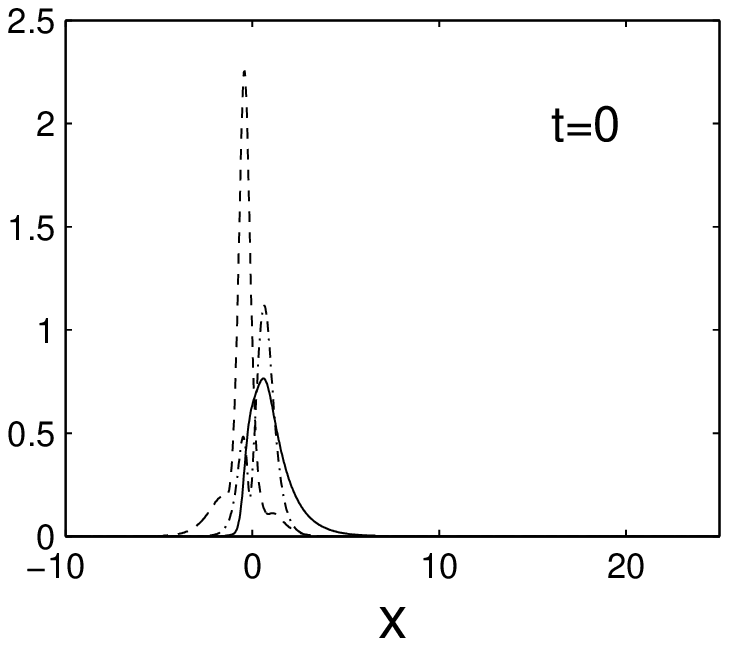}{1.0}} \hspace{0.4cm}
\parbox{5cm}{\postscript{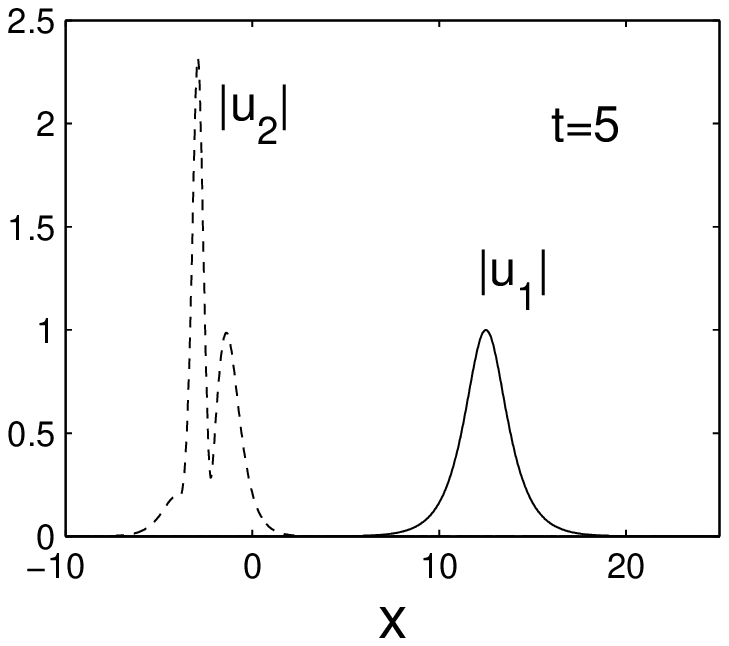}{1.0}}

\vspace{0.5cm} \caption{A non-generic higher-order soliton solution
which describes the $u_2+u_3 \to u_1+u_2$ process. The solution
parameters are the same as in Fig. 1 except that $\theta^{(1)}_1=0$
now. }
\end{center}
\end{figure}

\begin{figure}[p]
\begin{center}
\parbox{5cm}{\postscript{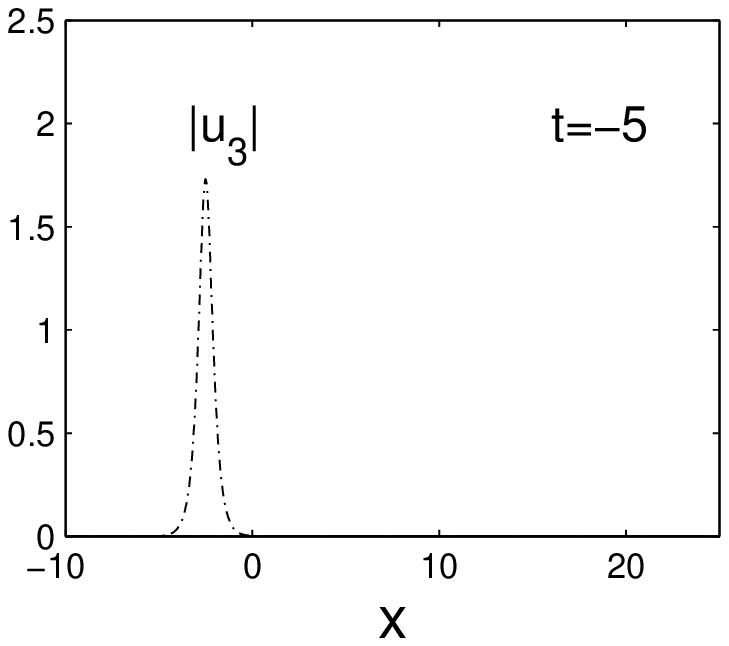}{1.0}} \hspace{0.4cm}
\parbox{5cm}{\postscript{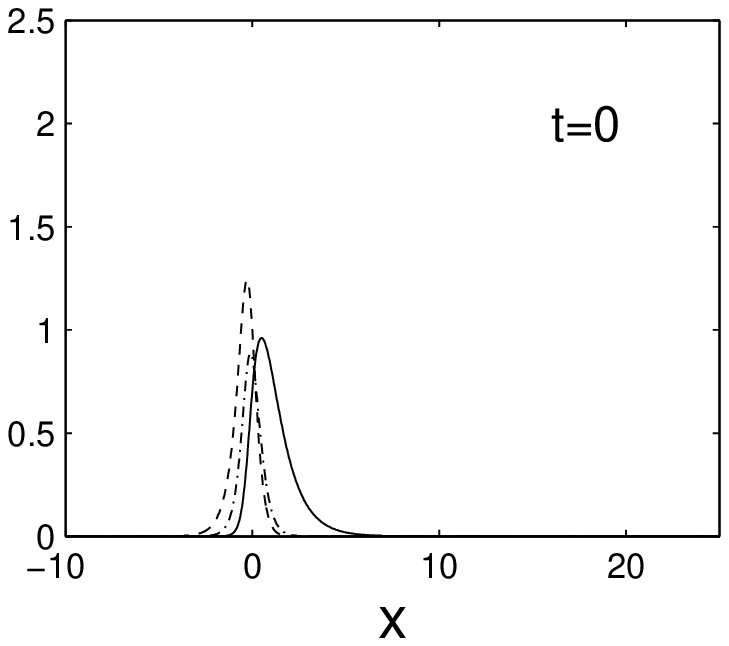}{1.0}} \hspace{0.4cm}
\parbox{5cm}{\postscript{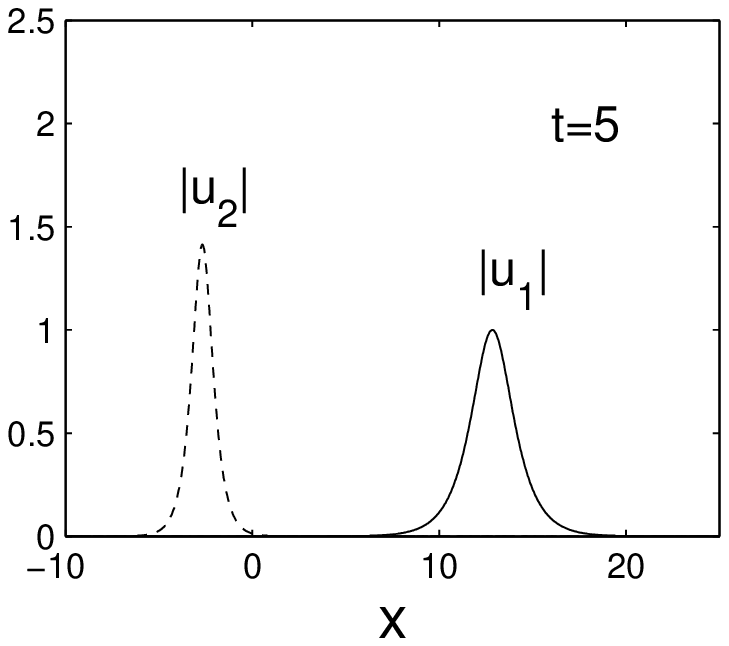}{1.0}}

\vspace{0.5cm} \caption{Another non-generic higher-order soliton
solution which describes the breaking of the $u_3$ sech wave into $u_1$
and $u_2$ sech waves. The solution parameters are the same as in Fig. 1
except that $\theta^{(1)}_1=\theta^{(1)}_2=0$ here.  }
\end{center}
\end{figure}

\begin{figure}[p]
\begin{center}
\parbox{5cm}{\postscript{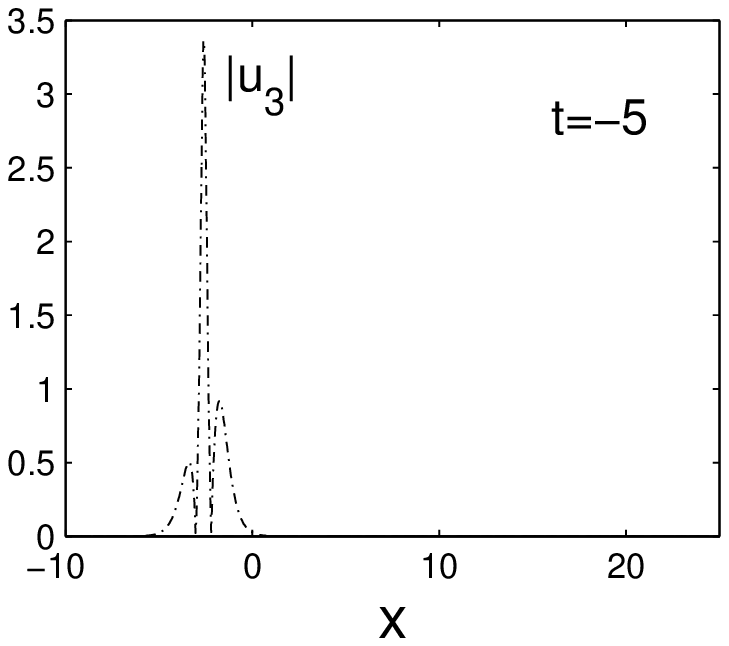}{1.0}} \hspace{0.4cm}
\parbox{5cm}{\postscript{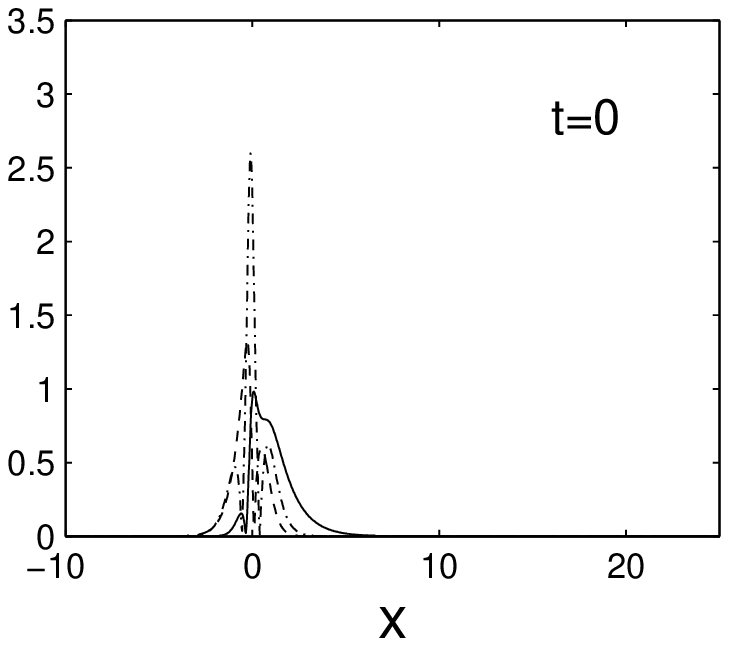}{1.0}} \hspace{0.4cm}
\parbox{5cm}{\postscript{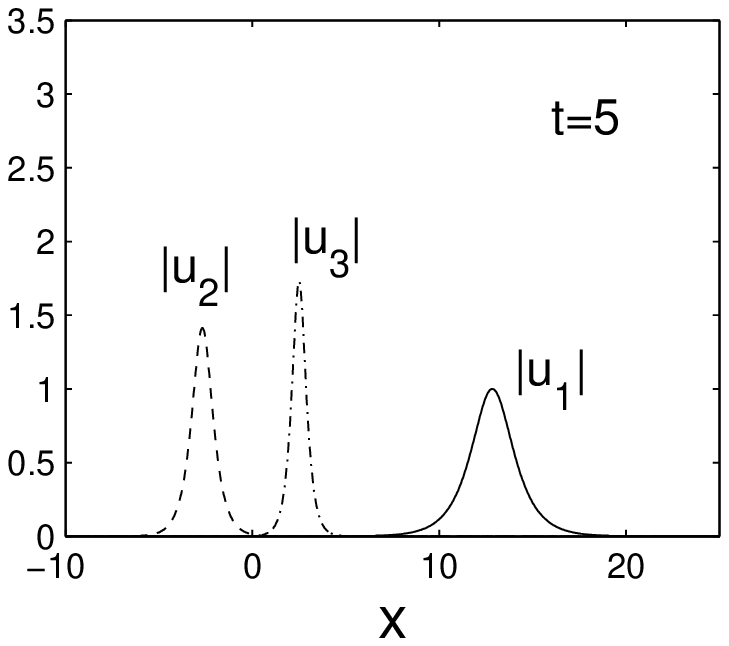}{1.0}}

\vspace{0.5cm} \caption{A non-generic higher-order soliton solution
which describes the $u_3 \to u_1+u_2+u_3$ process. The solution
parameters are the same as in Fig. 1 except that $\theta^{(1)}_2=0$
here. }
\end{center}
\end{figure}

\end{document}